\documentclass[12pt, onecolumn, draftclsnofoot, journal]{IEEEtran}

\usepackage[encapsulated]{CJK}
\usepackage{ucs}
\usepackage[utf8x]{inputenc}
\usepackage[cmex10]{amsmath}
\usepackage{amsmath,amssymb,amscd,bbm,amsthm,mathrsfs,dsfont}
\usepackage{algorithmic,algorithm}
\usepackage{mdwmath}
\usepackage{mdwtab}
\usepackage{bm,upgreek}
\usepackage{cite}
\usepackage{rotating,graphics,psfrag,epsfig}
\usepackage{array}
\usepackage{booktabs}
\usepackage{indentfirst}
\usepackage{subfigure}
\usepackage{lipsum,fancyhdr,lastpage,refcount}
\usepackage{url}
\usepackage{blindtext}
\usepackage[T1]{fontenc}
\usepackage{color}
\usepackage{stfloats}

\IEEEoverridecommandlockouts

\let\oldnl\nl
\newcommand{\nonl}{\renewcommand{\nl}{\let\nl\oldnl}}

\begin{document}

\title{Learning to Entangle Radio Resources in Vehicular Communications: An Oblivious Game-Theoretic Perspective}

\author{\IEEEauthorblockN{Xianfu Chen, Celimuge Wu, Mehdi Bennis, Zhifeng Zhao, and Zhu Han} 

\thanks{X. Chen is with the VTT Technical Research Centre of Finland, Oulu, Finland (email: xianfu.chen@vtt.fi). C. Wu is with the Graduate School of Informatics and Engineering, University of Electro-Communications, Tokyo, Japan (email: clmg@is.uec.ac.jp). M. Bennis is with the Centre for Wireless Communications, University of Oulu, Finland (email: mehdi.bennis@oulu.fi). Z. Zhao is with the College of Information Science and Electronic Engineering, Zhejiang University, Hangzhou, China (e-mail: zhaozf@zju.edu.cn). Z. Han is with the Department of Electrical and Computer Engineering as well as the Department of Computer Science, University of Houston, Houston, TX, USA (e-mail: zhan2@mail.uh.edu).}
} 



\maketitle

\begin{abstract}

This paper studies the problem of non-cooperative radio resource scheduling in a vehicle-to-vehicle communication network.
The technical challenges lie in high vehicle mobility and data traffic variations.
Over the discrete scheduling slots, each vehicle user equipment (VUE)-pair competes with other VUE-pairs in the coverage of a road side unit (RSU) for the limited frequency to transmit queued packets.
The frequency allocation at the beginning of each slot by the RSU is regulated following a sealed second-price auction. Each VUE-pair aims to optimize the expected long-term performance.
Such interactions among VUE-pairs are modelled as a stochastic game with a semi-continuous global network state space.
By defining a partitioned control policy, we transform the stochastic game into an equivalent game with a global queue state space of finite size.
We adopt an oblivious equilibrium (OE) to approximate the Markov perfect equilibrium (MPE), which characterizes the optimal solution to the equivalent game.
The OE solution is theoretically proven to be with an asymptotic Markov equilibrium property.
Due to the lack of a priori knowledge of network dynamics, we derive an online algorithm to learn the OE policies.
Numerical simulations validate the theoretical analysis and show the effectiveness of the proposed online learning algorithm.

\end{abstract}


\section{Introduction}
\label{intr}

The next generation vehicle-to-everything (V2X) technologies have been receiving increasing attentions for enabling emerging vehicular services, such as traffic safety, congestion reporting and in-vehicle infotainment \cite{Wu18, Camp17, Rost16}.
In particular, vehicle-to-vehicle (V2V) communication, operating in an ad hoc manner, provides more flexibility to render more attractive vehicle-related applications \cite{Kuut18}.
This type of vehicular applications have an ontological feature of requiring coordinations among the vehicles in close proximity \cite{Amad16}.
However, the topology of a V2V communication network changes dynamically across the time horizon because of the high vehicle mobility.
Without the support of an infrastructure, this in turn makes the design of radio resource management (RRM) techniques extremely challenging \cite{Zhen15}.

In the literature, there are a number of works focusing on RRM in V2V communications.
In \cite{Bai11}, Bai et al. proposed a low-complexity outage-optimal distributed channel allocation scheme for V2V communications based on maximum matching.
In \cite{Sun16}, Sun et al. investigated RRM for device-to-device based V2V communications, for which a separate resource block and power allocation algorithm was proposed.
Yao et al. proposed in \cite{Yao17} a loss differentiation rate adaptation scheme to meet the stringent delay and reliability requirements for V2V safety communications.
In \cite{Egea16}, Egea-Lopez et al. proposed a fair adaptive beaconing rate for the inter-vehicular communications algorithm to solve the problem of beaconing rate control.
Most of the efforts have ignored the network dynamics in the transmission quality as well as the data traffic variations, and hence fail to characterize the long-term RRM performance.

The framework of a Markov decision process (MDP) has been applied to formulate the problem of RRM in vehicular networks with time-varying nature.
In \cite{Liu18}, Liu et al. formulated the problem of power minimization with latency and reliability constraints and leveraged the Lyapunov stochastic optimization to deal with the network dynamics.
The same technique was adopted to study the problem of joint power and resource allocation for ultra reliable low latency communication in vehicular networks by Samarakoon et al. in \cite{Sama18}.
The Lyapunov stochastic optimization only constructs an approximately optimal solution.
In \cite{Zhen16}, Zheng et al. used a decentralized stochastic learning algorithm developed in \cite{Cui11} for the delay-aware radio resource scheduling in a software-defined vehicular network.
The proposed linear decomposition technique neglects the coupling of decision makings among the participating agents.
In \cite{Chen17}, we investigated the problem of non-cooperative RRM in a V2V communication network from an oblivious game-theoretic perspective and put forward an online learning algorithm to approach the solution.
However, our priori work does not take into account frequency resource sharing among different groups of vehicle user equipment (VUE)-pairs, which is highly dependent on the vehicular mobility characteristics.

In this paper, we are primarily concerned with the problem of non-cooperative radio resource scheduling in a V2V communication network.
Over an infinite discrete time horizon, each VUE-pair competes with other VUE-pairs in the coverage of a road side unit (RSU) for the limited frequency resource in order to transmit the queued data packets.
The frequency resource allocation at the beginning of each scheduling slot is centralized at the RSU and is regulated by a sealed second-price auction \cite{Vick61}\footnote{The dominant policy for a VUE-pair is to bid truthfully for the frequency resource.}.
The objective of a VUE-pair is to optimize the expected performance according to the network dynamics over a long-run.
Our major technical contributions from this work are summarized as follows.
\begin{itemize}
  \item We model the competitive interactions among the VUE-pairs as a stochastic game, where each VUE-pair aims to optimize its own expected long-term performance.
  \item We define a partitioned control policy profile, based on which the stochastic game with a semi-continuous global network state space can then be transformed into an equivalent game with a global queue state space of finite size.
      The optimal solution to the equivalent game can be characterized by a Markov perfect equilibrium (MPE).
  \item When the number of VUE-pairs in the V2V communication network is huge, we propose to approximate the MPE by an oblivious equilibrium (OE) \cite{Adla10, Wein05} to combat the curse of dimensionality in the global queue state space.
  \item We propose an online algorithm to learn the OE solution.
        The online learning algorithm requires no a priori statistical knowledge of the network dynamics.
\end{itemize}
To the best of our knowledge, this work is the first to introduce the concept of a OE for radio resource scheduling in V2V communications.

In following Section \ref{systMode}, we introduce the considered V2V communication network model and the assumptions made throughout this paper.
In Section \ref{probForm}, we formulate the non-cooperative radio resource scheduling among the VUE-pairs as a stochastic game, and an equivalent game is reformulated, the solution of which is characterized by a MPE under a partitioned control policy profile.
In Section \ref{probSolv}, we approximate the MPE by a OE and derive an online learning algorithm to find the OE solution.
In Section \ref{simuResu}, we evaluate the proposed algorithm through numerical simulations.
Finally, we draw the conclusions in Section \ref{conc}.

\section{System Model}
\label{systMode}

\begin{figure}[t]
  \centering
  \includegraphics[width=36pc]{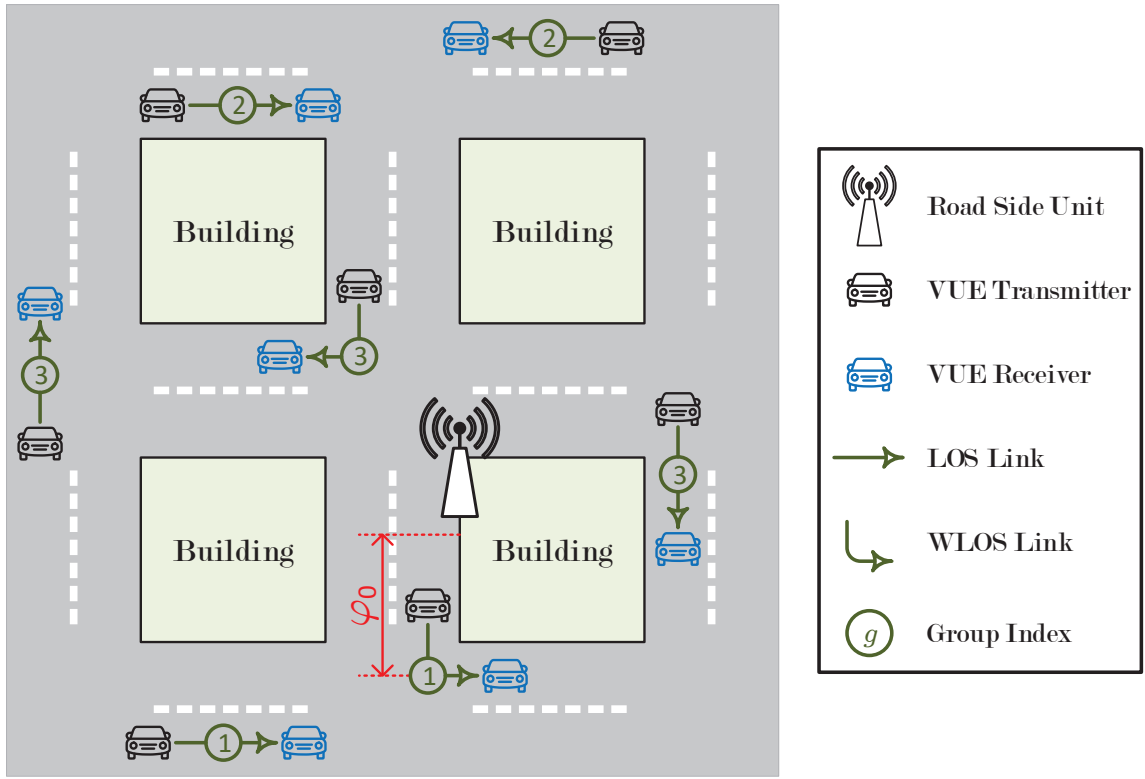}
  \caption{Illustration of a Manhattan grid vehicle-to-vehicle communication scenario (VUE: vehicle user equipment; LOS: line-of-sight; WLOS: weak-line-of-sight.).}
  \label{systModeFigu}
\end{figure}

As illustrated in Fig. \ref{systModeFigu}, this work considers a Manhattan grid V2V communication scenario, in which a set $\mathcal{K} = \{1, \cdots, K\}$ of VUE-pairs\footnote{It has been established that for a well defined road segment, the vehicle density approaches to be steady \cite{Zhua12}.} (each being associated with a VUE-transmitter (vTx) and a VUE-receiver (vRx)) compete for a common frequency resource within the coverage $\mathcal{L}$ of a RSU, where $\mathcal{L}$ represents a two-dimensional Euclidean space.
The whole system operates over the discrete scheduling slots, each of which is of equal duration $\delta$ and is indexed by a positive integer $t \in \mathds{N}_+$.
Let $\mathbf{x}_k^t = \left(x_{k}^{(1), t}, x_k^{(2), t}\right)$ and $\mathbf{y}_k^t = \left(y_k^{(1), t}, y_k^{(2), t}\right)$ be the Euclidean coordinates of the vTx and the vRx of a VUE-pair $k \in \mathcal{K}$ at slot $t$.
Over the time horizon, the VUE-pairs move in the coverage $\mathcal{L}$ according to a Manhattan mobility model \cite{Bai03}, and the vRxs always follow the vTxs with a fixed distance of $\varphi$.
In order to mitigate the interference during wireless transmissions and maximize the frequency utilization, the RSU clusters the VUE-pairs into a set $\mathcal{I}$ of disjoint groups based on their geographical locations, where $|\mathcal{I}| > 1$ with $|\mathcal{I}|$ denoting the cardinality of the set $\mathcal{I}$.
Depending on whether the vTx and the vRx of a VUE-pair $k$ are in the same lane or in perpendicular lanes, the channel model during each scheduling slot can be classified into: 1) line-of-sight (LOS) -- both the vTx and the vRx are in the same lane; 2) weak-line-of-sight (WLOS) -- the vTx and the vRx are in perpendicular lanes and at least one of them is near the intersection within a distance of $\varphi_0$; and otherwise, 3) none-line-of-sight (NLOS).
More specifically, the channel quality state $g_k^t = \nu_k^t \cdot H_k^t \in \mathcal{G}$ experienced by VUE-pair $k$ during scheduling slot $t$ includes a fast fading component $\nu_k^t$ of a Rayleigh distribution with a unit scale parameter and a path loss $H_k^t$ that applies the following model for urban areas using 5.9 GHz carrier frequency \cite{Abdu17},
\begin{align}\label{pathLossMode}
 H_k^t =
 \left\{\!\!
 \begin{array}{l@{~}l}
   \rho \cdot \left(\sqrt{\left|x_k^{(1), t} - x_k^{(2), t}\right|^2 + \left|y_k^{(1), t} - y_k^{(2), t}\right|^2}\right)^{-e},
                                                                                                        &\mbox{when VUE-pair $k$ in LOS};        \\
   \rho \cdot \left(\left|x_k^{(1), t} - x_k^{(2), t}\right| + \left|y_k^{(1), t} - y_k^{(2), t}\right|\right)^{-e},
                                                                                                        & \mbox{when VUE-pair $k$ in WLOS};      \\
   \xi  \cdot \left(\left|x_k^{(1), t} - x_k^{(2), t}\right| \cdot \left|y_k^{(1), t} - y_k^{(2), t}\right|\right)^{-e},
                                                                                                        & \mbox{when VUE-pair $k$ in NLOS},
 \end{array}
 \right.
\end{align}
where $e$ is the path loss coefficient while $\rho$ and $\xi$ are the path loss exponents with $\xi < \rho \cdot (\varphi_0 / 2)^e$.
As in \cite{Liu18}, we assume $\varphi_0 \geq \varphi$ for analytical tractability, which indicates $H_k^t$ depends on $\varphi$ only.
When $\varphi_0 < \varphi$, deep reinforcement learning \cite{Chen1801, Chen1802, Chen1803} can be adopted to address the explosion in global partitioned network states under a partitioned control policy profile, which will be defined later in Section \ref{equiGameRefo}.

During each scheduling slot, the RSU allocates the single frequency resource to the $|\mathcal{I}|$ groups, while in each group, we assume that the resource can be assigned to at most one vTx in order not to cause intra-group interference.
The centralized frequency resource allocation is regulated by the RSU using a sealed second-price auction.
Let $B_k^t$ be the bid submitted by each VUE-pair $k \in \mathcal{K}$ at the beginning of a scheduling slot $t$, and $\theta_k^t$ be an auction winner indicator that equals $1$ if VUE-pair $k$ wins the frequency resource and otherwise, $0$.
The winners are determined according to
\begin{align}\label{winnDete}
  \max_{\left\{\left(\theta_k^t \in \{0, 1\}: k \in \mathcal{K}\right): \sum_{k \in \mathcal{K}_i} \theta_k^t \leq 1, \forall i \in \mathcal{I}\right\}}
  \sum_{k \in \mathcal{K}} \theta_k^t \cdot B_k^t,
\end{align}
and the incurred payment to the RSU for each VUE-pair $k \in \mathcal{K}_i$ in a group $i \in \mathcal{I}$ at scheduling slot $t$ is calculated as
\begin{align}\label{paym}
  \tau_k^t = \theta_k^t \cdot \max_{k' \in \mathcal{K}_i \setminus \{k\}} B_{k'}^t,
\end{align}
which is resulted from the frequency access.

We assume that a data queue is maintained at the vTx of each VUE-pair $k \in \mathcal{K}$ to buffer the arriving data packets but may be terminated with a probability of $1 - \gamma \in (0, 1)$ after participating in the resource auction.
The data packet arrivals at the vTx constitutes a sequence of independent and random variables.
The winners from the resource auction acquire the right to access the frequency and proceed to transmit the queued data packets during the scheduling slot.
Let $q_k^t$ and $a_k^t$ be the queue length and the random new packet arrivals at scheduling slot $t$, respectively.
The queue evolution for VUE-pair $k$ can be expressed as
\begin{align}\label{queuEvol}
    q_k^{t + 1} =
    \left\{\!\!
    \begin{array}{l}
      0,                                                                        \mbox{ if the queue is terminated at slot } t;  \\
      \min\!\left\{q_k^t - \theta_k^t \cdot D_k^t + a_k^t, q^{(\max)}\right\},  \mbox{ otherwise},
    \end{array}
    \right.
\end{align}
where $q^{(\max)}$ is the maximum queue length such that $q_k^t \in \mathcal{Q} = \left\{0, \cdots, q^{(\max)}\right\}$ and $D_k^t$ is the scheduled number of packet departures during a scheduling slot $t$.
To simplify the wireless communication model, perfect channel state information is assumed.
The required transmit power for delivering $\theta_k^t \cdot D_k^t$ error-free data packets can be computed as
\begin{align}\label{poweCons}
  c_k^t = \dfrac{N + w \cdot \sigma^2}{g_k^t} \cdot \left(2^{\frac{\mu \cdot \theta_k^t \cdot D_k^t}{w \cdot \delta}} - 1\right),
\end{align}
where $N$ is the received aggregate interference due to inter-group frequency reuse, $w$ is the frequency bandwidth, $\sigma^2$ is the power spectral density of additive background noise, and $\mu$ is the constant size of a data packet.
Let $c^{(\max)}$ be the maximum transmit power for all vTxs, then $c_k^t \leq c^{(\max)}$, $\forall t$.

\section{Problem Description}
\label{probForm}

In this section, we formulate the problem of non-cooperative radio resource scheduling in the considered V2V network as a stochastic game and discuss the MPE solution.

\subsection{Network Dynamics}

During each scheduling slot $t$, the local state of a VUE-pair $k \in \mathcal{K}$ can be described by $\mathbf{s}_k^t = \left(g_k^t, (\mathbf{x}_k^t, \mathbf{y}_k^t), q_k^t\right) \in \mathcal{S} = \mathcal{G} \times \mathcal{L} \times \mathcal{Q}$, which includes the information of channel quality, geographical location and queue state, and $\mathbf{s}^t = \left(\mathbf{s}_k^t, \mathbf{s}_{-k}^t\right) \in \mathcal{S}^K$ is used to represent the global network state.
Herein, $-k$ denotes all the other VUE-pairs in set $\mathcal{K}$ without the presence of VUE-pair $k$.
For the urban Manhattan area, we assume that the VUE-pairs play a symmetric stationary control policy $\bm\pi = \left(\pi^{(\mathrm{f})}, \pi^{(\mathrm{p})}\right)$, which consists of the frequency resource auction policy $\pi^{(\mathrm{f})}$ and the packet scheduling policy $\pi^{(\mathrm{p})}$.
Note that $\pi^{(\mathrm{p})}$ is local network state dependent.
We let $\bm\pi_{(K)} = \left(\bm\pi_{(K)}^{(\mathrm{f})}, \bm\pi_{(K)}^{(\mathrm{p})}\right)$ denote the control policy vector where all $K$ VUE-pairs choose $\bm\pi$.
With $\bm\pi$, after observing the global network state $\mathbf{s}^t$ at the beginning of scheduling slot $t$, each VUE-pair $k$ makes decisions, namely, submitting a bid $B_k^t$ to the RSU for frequency resource allocation and scheduling the packet transmissions based on the auction results.
That is, $\bm\pi(\mathbf{s}^t) = \left(\pi^{(\mathrm{f})}(\mathbf{s}^t), \pi^{(\mathrm{p})}\left(\mathbf{s}_k^t\right)\right) = \left(B_k^t, D_k^t\right)$.
From the assumptions on the mobility of a VUE-pair, the queue evolution and the packet arrivals, the randomness lying in $\{\mathbf{s}^t: t \in \mathds{N}_+\}$ is Markovian with the following controlled state transition probability
\begin{align}\label{statTranProb}
 & \mathbb{P}\!\left(\mathbf{s}^{t + 1} | \mathbf{s}^t, \bm\theta^t\!\left(\bm\pi_{(K)}^{(\mathrm{f})}\left(\mathbf{s}^t\right)\right),
                       \bm\pi_{(K)}^{(\mathrm{p})}\left(\mathbf{s}^t\right)\right) =                                                            \\
 & \prod_{k \in \mathcal{K}} \mathbb{P}\!\left(g_k^{t + 1} | \left(\mathbf{x}_k^{t + 1}, \mathbf{y}_k^{t + 1}\right)\right) \cdot
     \mathbb{P}\!\left(\left(\mathbf{x}_k^{t + 1}, \mathbf{y}_k^{t + 1}\right) | \left(\mathbf{x}_k^t, \mathbf{y}_k^t\right)\right) \cdot 
     \mathbb{P}\!\left(q_k^{t + 1} | q_k^t, \theta_k^t\!\left(\bm\pi_{(K)}^{(\mathrm{f})}\left(\mathbf{s}^t\right)\right),                      \nonumber
                       \pi^{(\mathrm{p})}\left(\mathbf{s}_k^t\right)\right),
\end{align}
where $\mathbb{P}(\cdot)$ denotes the probability of an event and $\bm\theta^t = \left(\theta_k^t: k \in \mathcal{K}\right)$.

\subsection{Stochastic Game Formulation}

A payoff function is needed to reward a VUE-pair for winning the frequency resource auction.
The instantaneous payoff associated with each VUE-pair $k \in \mathcal{K}$ at each scheduling slot $t$ is chosen to be
\begin{align}\label{payoFunc}
  \ell_k\!\left(\mathbf{s}^t, \theta_k^t, D_k^t\right) = u_k\!\left(\mathbf{s}^t, \theta_k^t, D_k^t\right) - \tau_k^t,
\end{align}
where the utility function
\begin{align}\label{utilFunc}
  u_k\!\left(\mathbf{s}^t, \theta_k^t, D_k^t\right) =
  u_k^{(1)}\!\left(q_k^t\right) + \alpha_k u_k^{(2)}\!\left(c_k^t\right) + u_k^{(3)}\!\left(o_k^t\right),
\end{align}
with the packet overflows $o_k^t$ being given as follows
\begin{align}\label{packOve}
  o_k^t = \max\!\left\{q_k^t - \theta_k^t \cdot D_k^t + a_k^t - q^{(\max)}, 0\right\}.
\end{align}
Constrained by the finite buffer size at a vTx, the packet overflows occur when the arriving data packets cannot be all accepted to the queue.
In (\ref{utilFunc}), $\alpha_k > 0$ is a weight that trades off the importance of transmit power consumption, and $u_k^{(1)}(\cdot)$, $u_k^{(2)}(\cdot)$ and $u_k^{(3)}(\cdot)$ are the positive monotonically decreasing functions measuring the satisfactions of the queue length $q_k^t$ at the beginning of a scheduling slot, the transmit power consumption $c_k^t$ during a scheduling slot and the packet overflows $o_k^t$ in the end of a scheduling slot, respectively.

Due to the limited frequency resource and the stochastic nature in a V2V networking environment, we, therefore, formulate the problem of radio resource scheduling among the non-cooperative VUE-pairs over the infinite time-horizon as a stochastic game, in which $K$ VUE-pairs are the competitive players and there are a set $\mathcal{S}^K$ of global network states and a collection of stationary control policies $\bm\pi_{(K)}$.
We define a generic expected long-term payoff function $V_k(\mathbf{s} | \underline{\bm\pi}, \bm\pi_{(K - 1)})$ for a VUE-pair $k \in \mathcal{K}$ under the global network state $\mathbf{s}_k^t = \mathbf{s} =$ $(g_k, (\mathbf{x}_k, \mathbf{y}_k), q_k)$ at a current slot $t$, given that the other $K - 1$ competing VUE-pairs follow a common control policy $\bm\pi$ while VUE-pair $k$ follows $\underline{\bm\pi} = \left(\underline{\pi}^{(\mathrm{f})}, \underline{\pi}^{(\mathrm{p})}\right)$.
Specifically, we have
\begin{align}\label{statValu}
   V_k\!\left(\mathbf{s} | \underline{\bm\pi}, \bm\pi_{(K - 1)}\right) =
   \textsf{E}_{\left(\underline{\bm\pi}, \bm\pi_{(K - 1)}\right)}\!\!
   \left[\sum_{l = t}^{T_k^l} \ell_k\!\left(\mathbf{s}^l, \theta_k^l\!\left(\underline{\pi}^{(\mathrm{f})}\!\left(\mathbf{s}^l\right),
   \bm\pi_{(K - 1)}^{(\mathrm{f})}\!\left(\mathbf{s}^l\right)\right), \underline{\pi}^{(\mathrm{p})}\!\left(\mathbf{s}_k^l\right)\right) |
   \mathbf{s}^t = \mathbf{s}\right],
\end{align}
where $T_k^l \in \mathds{N}_+$ is the time that the data queue of VUE-pair $k$ terminates after scheduling slot $t$.
$V_k\left(\mathbf{s} | \underline{\bm\pi}, \bm\pi_{(K - 1)}\right)$ is also termed as the state value function of VUE-pair $k$ in a global network state $\mathbf{s}$ under a joint control policy $\left(\underline{\bm\pi}, \bm\pi_{(K - 1)}\right)$.
It can be found that $T_k^l$ is a geometric random variable.
We then equivalently express (\ref{statValu}) as
\begin{align}\label{statValuEqui}
 & V_k\!\left(\mathbf{s} | \underline{\bm\pi}, \bm\pi_{(K - 1)}\right) =                                                                        \nonumber\\
 & \textsf{E}_{\left(\underline{\bm\pi}, \bm\pi_{(K - 1)}\right)}\!\!
   \left[\sum_{l = t}^\infty (\gamma)^{l - t + 1} \cdot 
   \ell_k\!\left(\mathbf{s}^l, \theta_k^l\!\left(\underline{\pi}^{(\mathrm{f})}\!\left(\mathbf{s}^l\right),
   \bm\pi_{(K - 1)}^{(\mathrm{f})}\!\left(\mathbf{s}^l\right)\right), \underline{\pi}^{(\mathrm{p})}\!\left(\mathbf{s}_k^l\right)\right) |
   \mathbf{s}^t = \mathbf{s}\right],
\end{align}
where $(\gamma)^l$ denotes $\gamma$ to the $l$-th power.
Within a MDP framework, $\gamma$ can also be treated as a discount factor.
The aim of each VUE-pair $k$ is to find an optimal control policy $\bm\pi$ that maximizes $V_k\left(\mathbf{s} | \underline{\bm\pi}, \bm\pi_{(K - 1)}\right)$, $\forall \mathbf{s} \in \mathcal{S}^K$.

\subsection{Equivalent Game Reformulation}
\label{equiGameRefo}

From the channel model applied in this paper, the channel quality state space $\mathcal{G}$, where the path loss depends on the VUE mobility, is semi-continuous \cite{Bert87}.
Exploring the identical and independently distributed nature in channel quality states under a given path loss model, we utilize the notion of partitioned control policy below as in \cite{Wang13} to simply the non-cooperative stochastic game of radio resource scheduling.

\emph{Definition 1 (Partitioned Control Policy):}
Given a control policy $\bm\pi$, we define for any VUE-pair $k \in \mathcal{K}$,
\begin{align}\label{partContPoli}
  \bm\pi\!\left(\mathbf{q}\right) =
  \left\{\bm\pi(\mathbf{s}) | (g_k, (\mathbf{x}_k, \mathbf{y}_k)) \in \mathcal{G} \times \mathcal{L}, \forall k \in \mathcal{K}\right\},
\end{align}
as the collection of decision makings for all possible channel quality realizations given the global queue state $\mathbf{q} = (q_k, \mathbf{q}_{-k})$, where $q_k \in \mathcal{Q}$ is the local queue state at VUE-pair $k$ at a current scheduling slot.

With a partitioned control policy, we can turn the original MDP with a semi-continuous state space into a standard MDP with a finite state space.
The equivalent state value function, $V_k\left(\mathbf{q} | \underline{\bm\pi}, \bm\pi_{(K - 1)}\right)$, $\forall \mathbf{q} \in \mathcal{Q}^K$, of a VUE-pair $k \in \mathcal{K}$ can be given by
\begin{align}\label{statValuNew}
 & V_k\!\left(\mathbf{q} | \underline{\bm\pi}, \bm\pi_{(K - 1)}\right) =                                                                   \nonumber\\
 & \textsf{E}_{\left(\underline{\bm\pi}, \bm\pi_{(K - 1)}\right)}\!\!
   \left[\sum_{l = t}^\infty (\gamma)^{l - t + 1} \cdot
   \bar{\ell}_k\!\left(\mathbf{q}^l, \theta_k^l\!\left(\underline{\pi}^{(\mathrm{f})}\!\left(\mathbf{q}^l\right),
   \bm\pi_{(K - 1)}^{(\mathrm{f})}\!\left(\mathbf{q}^l\right)\right), \underline{\pi}^{(\mathrm{p})}\!\left(q_k^l\right)\right) |
   \mathbf{q}^t = \mathbf{q}\right],
\end{align}
where $\mathbf{q}^l = \left(q_k^l, \mathbf{q}_{-k}^l\right)$ is the global queue state at a scheduling slot $l$ and the corresponding instantaneous payoff function $\bar{\ell}_k\left(\mathbf{q}^l, \theta_k^l\left(\underline{\pi}^{(\mathrm{f})}\left(\mathbf{q}^l\right), \bm\pi_{(K - 1)}^{(\mathrm{f})}\left(\mathbf{q}^l\right)\right), \underline{\pi}^{(\mathrm{p})}\left(q_k^l\right)\right)$ is given by
\begin{align}\label{payoEqui}
 & \bar{\ell}_k\!\left(\mathbf{q}^l, \theta_k^l\!\left(\underline{\pi}^{(\mathrm{f})}\!\left(\mathbf{q}^l\right),
                       \bm\pi_{(K - 1)}^{(\mathrm{f})}\!\left(\mathbf{q}^l\right)\right), \underline{\pi}^{(\mathrm{p})}\!\left(q_k^l\right)\right) =  \nonumber\\
 & \textsf{E}_{\left\{\left(g_k^l, \left(\mathbf{x}_k^l, \mathbf{y}_k^l\right)\right): k \in \mathcal{K}\right\}}\!\!
    \left[\ell_k\!\left(\mathbf{s}^l, \theta_k^l\!\left(\underline{\pi}^{(\mathrm{f})}\!\left(\mathbf{s}^l\right),
           \bm\pi_{(K - 1)}^{(\mathrm{f})}\!\left(\mathbf{s}^l\right)\right), \underline{\pi}^{(\mathrm{p})}\!\left(\mathbf{s}_k^l\right)\right)\right].
\end{align}
Each VUE-pair $k$ hence switches to focus on designing an optimal stationary partitioned control policy $\bm\pi$ such that its own $V_k\left(\mathbf{q} | \underline{\bm\pi}, \bm\pi_{(K - 1)}\right)$ is maximized, $\forall \mathbf{q} \in \mathcal{Q}^K$.
In the equivalent radio resource scheduling stochastic game, a MPE defines the joint partitioned control policy profile $\bm\pi_{(K)}$ that simultaneously maximizes the expected long-term payoff for every VUE-pair in the network, given the partitioned control policies of the other VUE-pairs.

\emph{Definition 2 (Markov Perfect Equilibrium):}
The vector $\bm\pi_{(K)}$ of stationary partitioned control policies is a MPE in the equivalent radio resource scheduling stochastic game if $\forall k \in \mathcal{K}$ and $ \forall \mathbf{q} \in \mathcal{Q}^K$, we have
\begin{align}\label{MPE}
  V_k\!\left(\mathbf{q} | \bm\pi_{(K)}\right) = \max_{\underline{\bm\pi}} V_k\!\left(\mathbf{q} | \underline{\bm\pi}, \bm\pi_{(K - 1)}\right).
\end{align}

The following theorem ensures the existence of a MPE.

\emph{Theorem 1:}
In the equivalent non-cooperative radio resource scheduling game, there always exists a MPE \cite{Nowa92}.

\subsection{Solving the MPE}

The problem in (\ref{MPE}) is a typical infinite-horizon discounted MDP.
Suppose that each VUE-pair $k \in \mathcal{K}$ in the network can observe the global queue states and all the other VUE-pairs play $\bm\pi_{(K - 1)}$, the partitioned control policy $\bm\pi$ of VUE-pair $k$ satisfying (\ref{MPE}) can be obtained from solving the Bellman's equation,
\begin{align}\label{BellEqua}
 &    V_k\!\left(\mathbf{q} | \bm\pi_{(K)}\right)
   = \max_{\underline{\bm\pi}(\mathbf{s})}\!\bigg\{\gamma \cdot
     \bar{\ell}_k\!\left(\mathbf{q}, \theta_k\!\left(\underline{\pi}^{(\mathrm{f})}\!\left(\mathbf{q}\right),
     \bm\pi_{(K - 1)}^{(\mathrm{f})}\!\left(\mathbf{q}\right)\right), \underline{\pi}^{(\mathrm{p})}\!\left(q_k\right)\right)               \nonumber\\
 & + \gamma \cdot \sum_{\mathbf{q}' \in \mathcal{Q}^K} \mathbb{P}\!\left(\mathbf{q}' | \mathbf{q},
     \bm\theta\!\left(\underline{\pi}^{(\mathrm{f})}\!\left(\mathbf{q}\right),
     \bm\pi_{(K - 1)}^{(\mathrm{f})}\!\left(\mathbf{q}\right)\right), \left(\underline{\pi}^{(\mathrm{p})}\!\left(q_k\right),
     \bm\pi_{(K - 1)}^{(\mathrm{p})}\!\left(\mathbf{q}_{-k}\right)\right)\right) \cdot V_k\!\left(\mathbf{q}' | \bm\pi_{(K)}\right)\bigg\},
\end{align}
where $\bm\theta = (\theta_k: k \in \mathcal{K})$ is the result from the frequency resource auction at the beginning of a current scheduling slot, $\mathbf{q}' = (q_k', q_{-k}')$ is the subsequent global queue state, and the global queue state transition probability satisfies
\begin{align}\label{partStatTran}
 & \mathbb{P}\!\left(\mathbf{q}' | \mathbf{q}, \bm\theta\!\left(\underline{\pi}^{(\mathrm{f})}\!\left(\mathbf{q}\right),
     \bm\pi_{(K - 1)}^{(\mathrm{f})}\!\left(\mathbf{q}\right)\right), \left(\underline{\pi}^{(\mathrm{p})}\!\left(q_k\right),
     \bm\pi_{(K - 1)}^{(\mathrm{p})}\!\left(\mathbf{q}_{-k}\right)\right)\right) =                                                  \nonumber\\
 & \textsf{E}_{\left\{\left(g_k, \left(\mathbf{x}_k, \mathbf{y}_k\right)\right): k \in \mathcal{K}\right\}}\!\!
    \left[\mathbb{P}\!\left(\mathbf{q}' | \mathbf{s}, \bm\theta\!\left(\bm\pi_{(K)}^{(\mathrm{f})}\left(\mathbf{s}\right)\right),
                       \bm\pi_{(K)}^{(\mathrm{p})}\left(\mathbf{s}\right)\right)\right].
\end{align}
The solution to (\ref{BellEqua}) by a dynamic programming method \cite{Pute94} is in general computationally challenging.
The challenges lie in: 1) the number $\left(1 + q^{(\max)}\right)^K$ of global queue states, which grows exponentially as the number $K$ of VUE-pairs increases; and 2) the complete information of local queue state dynamics from the VUE-pairs, which is infeasible to exchange in our competitive networking environment.

\section{Learning the OE Policy}
\label{probSolv}

This section addresses the technical challenges in solving a MPE by an approximate OE and theoretically quantifies the error between a MPE solution and a OE solution.
Moreover, we propose an online learning algorithm to approach the OE control policy.

\subsection{Approximating MPE via OE}
\label{apprMPEOE}

The challenges in solving a MPE for the equivalent stochastic game motivates our alternative approach.
Basically, the idea is that in a dense network, there are a large number of VUE-pairs, and hence the impacts from competitions among the VUE-pairs on the frequency resource allocation can be averaged out such that the local queue states at the competitors remain approximately unchanged across the scheduling slots.
That is, as the number of VUE-pairs increases, the effect from a single VUE-pair on the outcomes of the equivalent stochastic game is negligible \cite{Adla10}.
Under this setting, each VUE-pair can potentially behave nearly optimally based only on the local queue states and the statistics of the long-term queue state distribution of other competing VUE-pairs.

\emph{Definition 3 (Long-Term Queue State Distribution):}
From the perspective of a VUE-pair $k \in \mathcal{K}$ playing $\underline{\bm\pi}$, the statistics of the long-term queue state distribution under $\left(\underline{\bm\pi}, \bm\pi_{(K - 1)}\right)$ is a mapping: $\Pi_k^{\left(\underline{\bm\pi}, \bm\pi_{(K - 1)}\right)}: \mathcal{S} \rightarrow [0, 1]$.
Specifically, $\forall q \in \mathcal{Q}$,
\begin{align}\label{PopuStat}
   \Pi_k^{\left(\underline{\bm\pi}, \bm\pi_{(K - 1)}\right)}(q) =
   \textsf{E}_{\left(\underline{\bm\pi}, \bm\pi_{(K - 1)}\right)}\!\! \left[\lim_{\Delta \rightarrow \infty}
   \frac{\sum_{l = t}^{t + \Delta} \sum_{k' \in \mathcal{K} \setminus \{k\}} \mathds{1}_{\left\{q_{k'}^l = q\right\}}}
   {(\Delta + 1) \cdot (K - 1)}\right],
\end{align}
where $\mathds{1}_{\{\Upsilon\}}$ is a function that equals $1$ if the condition $\Upsilon$ is satisfied and otherwise, $0$.

As stated in Lemma 2, the stationary behaviours from the VUE-pairs in the V2V network leads to a steady distribution over the global queue states.

\emph{Lemma 2:}
For the given stationary partitioned control policy profile $\left(\underline{\bm\pi}, \bm\pi_{(K - 1)}\right)$ following which a VUE-pair $k \in \mathcal{K}$ plays $\underline{\bm\pi}$ while all the other VUE-pairs play $\bm\pi_{(K - 1)}$, a steady long-term queue state distribution as in (\ref{PopuStat}) exists \cite{Gwon13}.

Therefore, in the following, we restrict a partitioned control policy $\underline{\bm\pi}$ for a VUE-pair $k \in \mathcal{K}$ to be oblivious.
Using an oblivious partitioned control policy, VUE-pair $k$ makes the frequency auction and packet scheduling decisions only with the local information.
Based on this intuition, we propose to approximate the MPE by a OE solution.
The corresponding oblivious state value function for VUE-pair $k$ can be defined as
\begin{align}\label{obviStat}
 & V_k\!\left(q_k | \underline{\bm\pi}, \Pi_k^{\left(\underline{\bm\pi}, \bm\pi_{(K - 1)}\right)}\right) =                                  \\
 & \textsf{E}_{\left(\underline{\bm\pi}, \bm\pi_{(K - 1)}\right)}\!\!
   \left[\sum_{l = t}^\infty (\gamma)^{l - t + 1} \cdot
   \bar{\ell}_k\!\left(\mathbf{q}^l, \theta_k^l\!\left(\underline{\pi}^{(\mathrm{f})}\!\left(q_k^l\right),
   \bm\pi_{(K - 1)}^{(\mathrm{f})}\!\left(\mathbf{q}_{-k}^l\right)\right), \underline{\pi}^{(\mathrm{p})}\!\left(q_k^l\right)\right) |
   q_k^t = q_k, \Pi_k^{\left(\underline{\bm\pi}, \bm\pi_{(K - 1)}\right)}\right],                                                           \nonumber
\end{align}
$\forall q_k \in \mathcal{Q}$.
Slightly different from the literature, we keep using the VUE-pair index $k$ for the sake of deriving an online learning algorithm to solve the OE control policy in Section \ref{learnOE}.

When all VUE-pairs play an optimal oblivious partitioned control policy profile $\bm\pi_{(K)}$, each VUE-pair $k \in \mathcal{K}$ conjectures $\Pi_k^{\bm\pi_{(K)}}$ as the statistics of the long-term queue state distribution that matches $\bm\pi_{(K)}$.
For notational convenience, we let $\Pi = \Pi_k^{\bm\pi_{(K)}}$, $\forall k \in \mathcal{K}$.
Then $\bm\pi_{(K)}$ together with $\Pi$ define a OE.

\emph{Definition 4 (Oblivious Equilibrium):}
A OE consists of a stationary oblivious partitioned control policy profile $\bm\pi_{(K)}$ and a long-term queue state distribution $\Pi$ such that $\forall k \in \mathcal{K}$,
\begin{align}\label{OE}
  V_k\!\left(q_k\right) =
  \max_{\underline{\bm\pi}} V_k\!\left(q_k | \underline{\bm\pi}, \Pi_k^{\left(\underline{\bm\pi}, \bm\pi_{(K - 1)}\right)}\right),
  \forall q_k \in \mathcal{Q},
\end{align}
where $V_k(q_k) = V_k(q_k | \bm\pi, \Pi)$.

In the equivalent radio resource scheduling stochastic game, a OE exists as in the following corollary.

\emph{Corollary 3:}
For the considered equivalent non-cooperative stochastic game of radio resource scheduling, the existence of a OE is straightforward from the discussions in \cite{Adla10}.

\subsection{Solution Error Analysis}

The aim in this subsection is to qualitatively analyze the error between a MPE and a OE solutions to the equivalent non-cooperative radio resource scheduling stochastic game.
We define an asymptotic Markov equilibrium (AME) property \cite{Weint08} for a OE control policy as follows.

\emph{Definition 5 (Asymptotic Markov Equilibrium Property):}
A OE control policy $\bm\pi_{(K)}$ is said to possess the AME property if for each VUE-pair $k \in \mathcal{K}$,
\begin{align}\label{asymMark}
  \lim_{K \rightarrow \infty}
  \textsf{E}\!\left[V_k\!\left(\mathbf{q} | \bm\pi_k, \bm\pi_{(K - 1)}\right) - V_k\!\left(\mathbf{q} | \bm\pi_{(K)}\right)\right] = 0,
\end{align}
where $\bm\pi_k$ is a MPE control policy of VUE-pair $k$.

The AME property means that the expected long-term payoff performance gap achieved by a VUE-pair $k \in \mathcal{K}$ from following a stationary OE control policy $\bm\pi$ instead of a MPE control policy $\bm\pi_k$ approaches zero when the number of VUE-pairs goes to infinity.
As a main result from this work, we verify in Theorem 4 the AME property of a OE.

\emph{Theorem 4:}
The AME property holds for the OE $(\bm\pi_{(K)}, \Pi)$.

\emph{Proof:}
For a control policy $\bm\pi_k = \left(\pi_k^{(\mathrm{f})}, \pi_k^{(\mathrm{p})}\right)$ played by a VUE-pair $k \in \mathcal{K}$, let us define, $\forall \mathbf{q} \in \mathcal{Q}^K$,
\begin{align}\label{diff}
  \Delta V_k(\mathbf{q}) = V_k\!\left(\mathbf{q} | \bm\pi_k, \bm\pi_{(K - 1)}\right) - V_k\!\left(\mathbf{q} | \bm\pi_{(K)}\right).
\end{align}
Without loss of generality, let $\Delta V_k(\mathbf{q}) \geq 0$.
The case in which $\Delta V_k(\mathbf{q}) < 0$ proceeds in a similar way.
We will establish that $\lim_{K \rightarrow \infty} \textsf{E}[\Delta V_k(\mathbf{q})] = 0$.
$\forall q_k \in \mathcal{Q}$, define the expected long-term payoff for VUE-pair $k$ by
\begin{align}\label{statValuevsObli}
 & V_k\!\left(q_k | \underline{\bm\pi}_k, \Pi_k^{\left(\underline{\bm\pi}_k, \bm\pi_{(K - 1)}\right)}\right) =                                  \\
 & \textsf{E}_{\left(\underline{\bm\pi}_k, \bm\pi_{(K - 1)}\right)}\!\!
   \left[\sum_{l = t}^\infty (\gamma)^{l - t + 1} \cdot
   \bar{\ell}_k\!\left(\mathbf{q}^l, \theta_k^l\!\left(\underline{\pi}_k^{(\mathrm{f})}\!\left(\mathbf{q}^l\right),
   \bm\pi_{(K - 1)}^{(\mathrm{f})}\!\left(\mathbf{q}_{-k}^l\right)\right), \underline{\pi}_k^{(\mathrm{p})}\!\left(q_k^l\right)\right) |
   q_k^t = q_k, \Pi_k^{\left(\underline{\bm\pi}_k, \bm\pi_{(K - 1)}\right)}\right]\!,                                                           \nonumber
\end{align}
under any stationary control policy $\underline{\bm\pi}_k= \left(\underline{\pi}_k^{(\mathrm{f})}, \underline{\pi}_k^{(\mathrm{p})}\right)$ while other VUE-pairs playing the oblivious $\bm\pi_{(K - 1)}$.
Then, $\forall q_k \in \mathcal{Q}$,
\begin{align}
   V_k\!\left(q_k | \bm\pi_k, \Pi_k^{\left(\bm\pi_k, \bm\pi_{(K - 1)}\right)}\right)
 & \leq
   \max_{\underline{\bm\pi}_k} V_k\!\left(q_k | \underline{\bm\pi}_k, \Pi_k^{\left(\underline{\bm\pi}_k, \bm\pi_{(K - 1)}\right)}\right) \nonumber\\
 & \overset{\mbox{(a)}}{=}
   \max_{\underline{\bm\pi}} V_k\!\left(q_k | \underline{\bm\pi}, \Pi_k^{\left(\underline{\bm\pi}, \bm\pi_{(K - 1)}\right)}\right)       \nonumber\\
 & = V_k(q_k),
\end{align}
where (a) is from \cite[Theorem 5.1]{Wein05}.
$\Delta V_k(\mathbf{q})$ in (\ref{diff}) can be written as
\begin{align}
     \Delta V_k(\mathbf{q})
 & = V_k\!\left(\mathbf{q} | \bm\pi_k, \bm\pi_{(K - 1)}\right) - V_k(q_k)
   + V_k(q_k) - V_k\!\left(\mathbf{q} | \bm\pi_{(K)}\right)                                                                \nonumber\\
 & \leq
     \underbrace{V_k\!\left(\mathbf{q} | \bm\pi_k, \bm\pi_{(K - 1)}\right)
   - V_k\!\left(q_k | \bm\pi_k, \Pi_k^{\left(\bm\pi_k, \bm\pi_{(K - 1)}\right)}\right)}_{\mathds{V}_k^{(1)}(\mathbf{q})}
   + \underbrace{V_k(q_k) - V_k\!\left(\mathbf{q} | \bm\pi_{(K)}\right)}_{\mathds{V}_k^{(2)}(\mathbf{q})}.
\end{align}

Using the triangle inequality, $\forall k \in \mathcal{K}$ and $\forall \mathbf{q} \in \mathcal{Q}^K$, we deduce
\begin{align}\label{V1}
 & \textsf{E}\!\left[\mathds{V}_k^{(1)}(\mathbf{q})\right] \leq                                                                     \\
 & \textsf{E}_{\left(\bm\pi_k, \bm\pi_{(K - 1)}\right)}\!\!\!
   \left[\sum_{l = t}^\infty (\gamma)^{l - t + 1} \!\cdot\!
   \left|\!\!
   \begin{array}{l}
      \bar{\ell}_k\!\left(\mathbf{q}^l, \theta_k^l\!\left(\pi_k^{(\mathrm{f})}\!\left(\mathbf{q}^l\right),
         \bm\pi_{(K - 1)}^{(\mathrm{f})}\!\left(\mathbf{q}^l\right)\right), \pi_k^{(\mathrm{p})}\!\left(q_k^l\right)\right) -       \\
      \bar{\ell}_k\!\left(\mathbf{q}^l, \theta_k^l\!\left(\pi_k^{(\mathrm{f})}\!\left(q_k^l\right),
         \bm\pi_{(K - 1)}^{(\mathrm{f})}\!\left(\mathbf{q}_{-k}^l\right)\right), \pi_k^{(\mathrm{p})}\!\left(q_k^l\right)\right)
   \end{array}\!\!
   \right|\! |
   \mathbf{q}^t = \mathbf{q}, \Pi_k^{\left(\bm\pi_k, \bm\pi_{(K - 1)}\right)}\right]\!\!,                                               \nonumber
\end{align}
and
\begin{align}\label{V2}
 & \textsf{E}\!\left[\mathds{V}_k^{(2)}(\mathbf{q})\right] \leq                                                                     \nonumber\\
 & \textsf{E}_{\bm\pi_{(K)}}\!\!
   \left[\sum_{l = t}^\infty (\gamma)^{l - t + 1} \!\cdot\!
   \left|\!\!
   \begin{array}{l}
      \bar{\ell}_k\!\left(\mathbf{q}^l, \theta_k^l\!\left(\pi^{(\mathrm{f})}\!\left(q_k^l\right),
         \bm\pi_{(K - 1)}^{(\mathrm{f})}\!\left(\mathbf{q}_{-k}^l\right)\right), \pi^{(\mathrm{p})}\!\left(q_k^l\right)\right) -    \\
      \bar{\ell}_k\!\left(\mathbf{q}^l, \theta_k^l\!\left(\pi_k^{(\mathrm{f})}\!\left(\mathbf{q}^l\right),
         \bm\pi_{(K - 1)}^{(\mathrm{f})}\!\left(\mathbf{q}^l\right)\right), \pi_k^{(\mathrm{p})}\!\left(q_k^l\right)\right)
   \end{array}\!\!
   \right| |
   \mathbf{q}^t = \mathbf{q}, \Pi\right].
\end{align}
\cite[Lemma 6]{Adla10} implies that both $\textsf{E}[\mathds{V}_k^{(1)}(\mathbf{q})]$ and $\textsf{E}[\mathds{V}_k^{(2)}(\mathbf{q})]$ approach to zero as $K \rightarrow \infty$, which completes the proof.
\hfill$\Box$

\subsection{Learning the OE}
\label{learnOE}

The Bellman's optimality equation for the problem in (\ref{OE}) can be written as
\begin{align}\label{Bell}
     V_k(q_k)
 & = \max_{\underline{\bm\pi}(q_k)} \bigg\{
     \gamma \cdot \bar{\ell}_k\!\left(\mathbf{q}, \theta_k\!\left(\underline{\pi}^{(\mathrm{f})}\!\left(q_k\right),
     \bm\pi_{(K - 1)}^{(\mathrm{f})}\!\left(\mathbf{q}_{-k}\right)\right), \underline{\pi}^{(\mathrm{p})}(q_k)\right)                     \nonumber\\
 & + \gamma \cdot \sum_{q_k' \in \mathcal{Q}} \mathbb{P}\!\left(q_k' | q_k,
     \theta_k\!\left(\underline{\pi}^{(\mathrm{f})}\!\left(q_k\right), \bm\pi_{(K - 1)}^{(\mathrm{f})}\!\left(\mathbf{q}_{-k}\right)\right),
     \underline{\pi}^{(\mathrm{p})}(q_k)\right) \cdot V_k(q_k')\bigg\},
\end{align}
$\forall k \in \mathcal{K}$ and $\forall q_k \in \mathcal{Q}$.
After observing the local queue state $q_k$ at the beginning of every scheduling slot, each VUE-pair $k \in \mathcal{K}$ strategically submits an auction bid $B_k = \pi^{(\mathrm{f})}(q_k)$ to the RSU, which is a true value of owning the single frequency resource.
The following Theorem 5 gives the optimal design of $B_k$.

\emph{Theorem 5:}
When all VUE-pairs follow a joint partitioned control policy profile $\bm\pi_{(K)}$, the optimal auction bid $B_k$ submitted by a VUE-pair $k \in \mathcal{K}$ at a current scheduling slot is of the form as
\begin{align}\label{auctBid}
 B_k = u_k\!\left(\mathbf{s}, \theta_k, D_k\right) + \sum_{q_k' \in \mathcal{Q}} \mathbb{P}\!\left(q_k' | q_k, \theta_k, D_k\right)
       \cdot V_k\!\left(q_k'\right),
\end{align}
where $\theta_k$ is the optimal frequency resource allocation from the auction at the RSU and $D_k = \pi^{(\mathrm{p})}(\mathbf{s}_k)$ is the optimal packet scheduling decision at a current scheduling slot.

\emph{Proof:}
Since a stationary oblivious control policy $\bm\pi$ is composed of the oblivious frequency resource auction policy $\pi^{(\mathrm{f})}$ and the oblivious packet scheduling policy $\pi^{(\mathrm{p})}$, the Bellman's optimality equation in (\ref{Bell}) for each VUE-pair $k \in \mathcal{K}$ can be restructured as 
\begin{align}\label{BellEqui}
     B_k
   = \pi^{(\mathrm{f})}(q_k)
 & = \underset{\underline{B}_k}{\arg\max}~
     \textsf{E}_{\left\{\left(g_k, \left(\mathbf{x}_k, \mathbf{y}_k\right)\right): k \in \mathcal{K}\right\}}\bigg\{
     \ell_k\!\left(\mathbf{s}, \theta_k\!\left(\underline{B}_k, \mathbf{B}_{-k}\right), \pi^{(\mathrm{p})}(\mathbf{s}_k)\right)         \nonumber\\
 & + \sum_{q_k' \in \mathcal{Q}}
     \mathbb{P}\!\left(q_k' | q_k, \theta_k\!\left(\underline{B}_k, \mathbf{B}_{-k}\right), \pi^{(\mathrm{p})}(\mathbf{s}_k)\right)
     \cdot V_k(q_k')\bigg\},
\end{align}
where $\mathbf{B}_{-k} = (B_{k'}: k' \in \mathcal{K} \setminus \{k\}) = \bm\pi_{(K - 1)}(\mathbf{s}_{-k})$ denotes the optimal frequency auction bids from other VUE-pairs.
From the winner determination (\ref{winnDete}) and the payment calculation (\ref{paym}) rules, the optimal oblivious frequency auction policy for VUE-pair $k$ is to bid truthfully across the scheduling slots with the bid defined by (\ref{auctBid}).
\hfill$\Box$

Directly calculating an auction bid at the beginning of a scheduling slot as in (\ref{auctBid}) remains challenging due to the facts that the packet arrival statistics may not be easily known a priori.
We define a post-decision queue state \cite{Mast13, Salo08, Chen1801} for each VUE-pair based on the observation that the packet arrivals are independent of the frequency resource auction and the packet scheduling decision makings.
At a current slot, the post-decision queue state of a VUE-pair $k \in \mathcal{K}$ is defined as $\tilde{q}_k = q_k - \theta_k(\mathbf{B}) \cdot D_k$, where $\mathbf{B} = (B_k: k \in \mathcal{K})$.
By introducing a post-decision queue state, we hence factor the utility function given by (\ref{utilFunc}) into two parts, namely, $u_k^{(1)}(\cdot) + \alpha_k \cdot u_k^{(2)}(\cdot)$ and $u_k^{(3)}(\cdot)$.
The probability of the local queue state transition from $q_k$ to $q_k'$ can be then expressed as
\begin{align}\label{locaPartStatTran}
     \mathbb{P}\!\left(q_k' | q_k, \theta_k(\mathbf{B}), D_k\right)
 & = \mathbb{P}\!\left(q_k' | \tilde{q}_k\right) \cdot
     \mathbb{P}\!\left(\tilde{q}_k | q_k, \theta_k(\mathbf{B}), D_k\right)      \nonumber\\
 & = \gamma \cdot \mathbb{P}\!\left(a_k\right) + (1 - \gamma),
\end{align}
where it is obvious that $\mathbb{P}(\tilde{q}_k | q_k, \theta_k(\mathbf{B}), D_k) = 1$.
We denote the optimal oblivious post-decision queue state value function of VUE-pair $k$ by $\tilde{V}_k(\tilde{q}_k)$, which can be expressed as
\begin{align}\label{postDeci0}
   \tilde{V}_k\!\left(\tilde{q}_k\right)
 = \gamma \cdot u_k^{(3)}(o_k) + \gamma \cdot \sum_{q_k' \in \mathcal{Q}} \mathbb{P}\!\left(q_k' | \tilde{q}_k\right) \cdot V_k(q_k'),
\end{align}
where $o_k$ is the number of incurred packet overflows.

For each VUE-pair $k \in \mathcal{K}$, we further define the right-hand-side of (\ref{Bell}) as a $Q$-factor, that is,
\begin{align}\label{QFact}
     Q_k\!\left(q_k, \theta_k(\underline{B}_k, \mathbf{B}_{-k}), \underline{D}_k\right)
 & = \gamma \cdot \bar{\ell}_k\!\left(\mathbf{q}, \theta_k(\underline{B}_k, \mathbf{B}_{-k}), \underline{D}_k\right)                     \nonumber\\
 & + \gamma \cdot \sum_{q_k' \in \mathcal{Q}} \mathbb{P}\!\left(q_k' | q_k,
     \theta_k(\underline{B}_k, \mathbf{B}_{-k}), \underline{D}_k\right) \cdot V_k(q_k'),
\end{align}
where $\underline{B}_k$ and $\underline{D}_k$ are, respectively, the resource auction bid and the packet scheduling decision in the local queue state $q_k$ at a current scheduling slot following an any given partitioned control policy $\underline{\bm\pi}$.
Then the optimal oblivious queue state value function $V_k(q_k)$ can be derived from
\begin{align}\label{ObliStatValu}
  V_k(q_k) = \max_{\underline{B}_k, \underline{D}_k} Q_k\!\left(q_k, \theta_k(\underline{B}_k, \mathbf{B}_{-k}), \underline{D}_k\right).
\end{align}
Replacing $V_k(q_k')$ in (\ref{postDeci0}) with (\ref{ObliStatValu}), we obtain
\begin{align}\label{postDeci}
     \tilde{V}_k\!\left(\tilde{q}_k\right)
   = \gamma \cdot u_k^{(3)}(o_k)
   + \gamma \cdot \sum_{q_k' \in \mathcal{Q}} \mathbb{P}\!\left(q_k' | \tilde{q}_k\right)
     \cdot \max_{\underline{B}_k', \underline{D}_k'} Q_k\!\left(q_k', \theta_k(\underline{B}_k', \mathbf{B}_{-k}), \underline{D}_k'\right).
\end{align}
And in turn, the $Q$-factor defined by (\ref{QFact}) can be given by
\begin{align}\label{QFact1}
   Q_k\!\left(q_k, \theta_k(\underline{B}_k, \mathbf{B}_{-k}), \underline{D}_k\right) =
   \gamma \cdot \left(u_k^{(1)}(q_k) + \alpha_k \cdot u_k^{(2)}\!\left(\underline{c}_k\right) -
   \underline{\tau}_k\right) + \tilde{V}_k(\tilde{q}_k),
\end{align}
where $\underline{\tau}_k$ is the induced payment for VUE-pair $k$ to the RSU from submitting a bid $\underline{B}_k$ and $\underline{c}_k$ is the power consumption for transmitting $\theta_k\left(\underline{B}_k, \mathbf{B}_{-k}\right) \cdot \underline{D}_k$ data packets.

By substituting (\ref{locaPartStatTran}) and (\ref{postDeci}) back into (\ref{auctBid}), we eventually arrive at the auction bid for a VUE-pair $k \in \mathcal{K}$,
\begin{align}\label{auctBidNew}
  B_k = u_k^{(1)}(q_k) + \alpha_k \cdot u_k^{(2)}(c_k) + \frac{1}{\gamma} \cdot \tilde{V}_k\!\left(\tilde{q}_k\right),
\end{align}
where the transmit power $c_k$ is consumed by VUE-pair $k$ to deliver a number $\theta_k\left(\bm\pi^{(\mathrm{f})}(\mathbf{q})\right) \cdot \pi^{(\mathrm{p})}(s_k)$ of data packets.

The number of packet arrivals by the end of a scheduling slot is unavailable beforehand and so is the number of packet overflows at the slot.
In this case, instead of computing the optimal oblivious post-decision state value function using (\ref{postDeci}), we propose an online algorithm for each VUE-pair $k \in \mathcal{K}$ to find the optimal oblivious post-decision queue state value function by exploring the conventional reinforcement learning techniques \cite{Salo08, Rich98}.
Based on the observations of local queue state $q_k^t$, frequency allocation result $\theta_k^t$ from the auction at the RSU, number of packet departures $\theta_k^t \cdot D_k^t$, consumed transmit power $c_k^t$, local post-decision queue state $q_k^t - \theta_k^t \cdot D_k^t$, number of new packet arrivals $a_k^t$, number of packet overflows $o_k^t$, payment $\tau_k^t$ to the RSU at current scheduling slot $t$, and resulting local queue state $q_k^{t + 1}$ at the next scheduling slot $t + 1$, VUE-pair $k$ updates the oblivious post-decision queue state value function on the fly according to
\begin{align}\label{LearPostDeci}
     \tilde{V}_k^{t + 1}\!\left(\tilde{q}_k^t\right)
   = \left(1 - \zeta^t\right) \cdot \tilde{V}_k^t\!\left(\tilde{q}_k^t\right)
   + \zeta^t \cdot \gamma \cdot \left(u_k^{(3)}\!\left(o_k^t\right)
   + \max_{\underline{\theta}_k', \underline{D}_k'} Q_k^t\!\left(q_k^{t + 1}, \underline{\theta}_k', \underline{D}_k'\right)\right).
\end{align}
where $\zeta^t \in [0, 1)$ is the learning rate, the $Q$-factor is iterated following the rule below
\begin{align}\label{QFactUpda}
     Q_k^{t + 1}\!\left(q_k^t, \theta_k^t, D_k^t\right)
  = \gamma \cdot \left(u_k^{(1)}\!\left(q_k^t\right) + \alpha_k \cdot u_k^{(2)}\!\left(c_k^t\right) - \tau_k^t\right)
  + \tilde{V}_k^{t + 1}\!\left(\tilde{q}_k^t\right),
\end{align}
and the number $D_k^t$ of scheduled packet departures during slot $t$ is determined by
\begin{align}\label{packDepa}
   D_k^t = \underset{\underline{D}_k}{\arg\max}~Q_k^t\!\left(q_k^t, \theta_k^t, \underline{D}_k\right).
\end{align}

The online algorithm for learning the optimal oblivious post-decision queue state value functions for each VUE-pair $k \in \mathcal{K}$ in the V2V network is briefly summarized in Algorithm \ref{algo1}.
\begin{algorithm}[t]
    \caption{Online Algorithm for Learning Optimal Oblivious Post-decision State Value Functions of a VUE-pair $k \in \mathcal{K}$}
    \label{algo1}
    \begin{algorithmic}[1]
        \STATE \textbf{initialize} the oblivious post-decision state value function $\tilde{V}_k^1(\tilde{q}_k)$ and $Q$-factor $Q_k^1(q_k, \theta_k, D_k)$ for VUE-pair $k$, where $\tilde{q}_k$, $q_k \in \mathcal{Q}$ and $(\theta_k, D_k) \in \{0, 1\} \times \mathcal{Q}$.

        \REPEAT
            \STATE At the beginning of each scheduling slot $t$, VUE-pair $k$ observes the local queue state $q_k^t$, calculates the auction bid $B_k^t$ according to (\ref{auctBidNew}), and sends the information of $\left[\left(\mathbf{x}_k^t, \mathbf{y}_k^t\right), B_k^t\right]$ to the RSU, where $D_k^t$ is determined according to (\ref{packDepa}).

            \STATE VUE-pair $k$ awaits the frequency resource allocation $\theta_k^t$ and the payment $\tau_k^t$, which are calculated according to (\ref{winnDete}) and (\ref{paym}), respectively. Then the vTx of VUE-pair $k$ proceed to make packet scheduling decision $D_k^t$.

            \STATE After transmitting $\theta_k^t \cdot D_k^t$ packets, VUE-pair $k$ observes the utilities $u_k^{(1)}(q_k^t)$ and $u_k^{(2)}(c_k^t)$ regarding the queue length $q_k^t$ and the transmit power consumption $c_k^t$, respectively, and the resulting local post-decision queue state $\tilde{q}_k^t = q_k^t - \theta_k^t \cdot D_k^t$.

            \STATE With the observation of $a_k^t$ new packet arrivals, VUE-pair $k$ realizes $u_k^{(3)}(o_k^t)$ quantifying the satisfaction of packet overflows at scheduling slot $t$ and the local queue state transits to $q_k^{t + 1} = \min\left\{\tilde{q}_k^t + a_k^t, q^{(\max)}\right\}$ during the following scheduling slot $t + 1$.

            \STATE According to (\ref{LearPostDeci}) and (\ref{QFactUpda}), VUE-pair $k$ updates, respectively, the oblivious post-decision queue state value function $\tilde{V}_k^{t + 1}(\tilde{q}_k^t)$ and the $Q$-factor $Q_k^{t + 1}(q_k^t, \theta_k^t, D_k^t)$.

            \STATE The scheduling slot index is updated by $t \leftarrow t+1$.
        \UNTIL{A predefined stopping condition is satisfied.}
    \end{algorithmic}
\end{algorithm}
Theorem 6 ensures the convergence property of the online learning algorithm.

\emph{Theorem 6:}
For each VUE-pair $k \in \mathcal{K}$, $\left\{\tilde{V}_k^t(\tilde{q}_k): \forall t\in\mathbb{N}_+\right\}$ converges to the optimal oblivious post-decision queue state value function $\tilde{V}_k(\tilde{q}_k)$, $\forall \tilde{q}_k \in \mathcal{Q}$, if and only if the learning rate satisfies $\sum_{t = 1}^\infty \zeta^t = \infty$ and $\sum_{t = 1}^\infty (\zeta^t)^2 < \infty$.

\emph{Proof:}
The proof is similar to \cite{Chen1804}.
\hfill$\Box$

\begin{figure}[t]
  \centering
  \includegraphics[width=29pc]{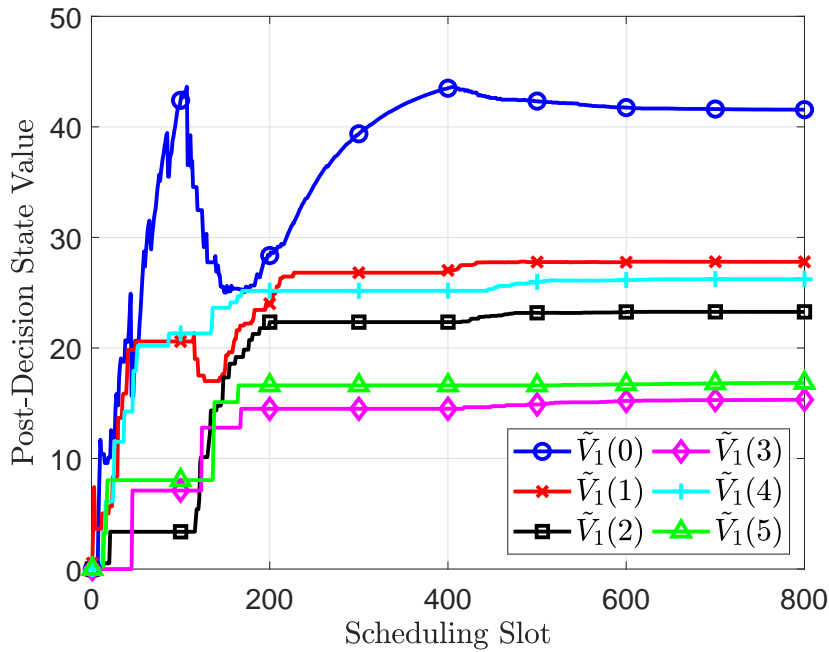}
  \caption{Illustration of the convergence property of our proposed algorithm.}
  \label{sim01}
\end{figure}

\section{Simulation Results}
\label{simuResu}

We carry out numerical simulations to evaluate the performance achieved from our proposed online learning algorithm for radio resource scheduling in a non-cooperative V2V communication network, which is based on a $250\times250$ m$^2$ Manhattan mobility model with nine intersections as in \cite{Liu18}.
In the model, a road consists of two lanes, each of which is in one direction and is of width $4$ m.
The average vehicle speed is $40$ km/h, and the vehicle grouping is performed by means of spectral clustering \cite{Luxb07}.
$u_k^{(1)}(\cdot)$, $u_k^{(2)}(\cdot)$ and $u_k^{(3)}(\cdot)$ in (\ref{utilFunc}) are chosen to be
\begin{align}
    u_k^{(1)}\!\left(q_k^t\right)  & = \exp\!\left\{-q_k^t\right\}, \\
    u_k^{(2)}\!\left(c_k^t\right)  & = \exp\!\left\{-c_k^t\right\}, \\
    u_k^{(3)}\!\left(o_k^t\right)  &= \exp\!\left\{-o_k^t\right\}.
\end{align}
The packet arrivals at the vTx of a VUE-pair follow a Poisson arrival process with average rate $\lambda$ (in packets per scheduling slot).
Other parameter values used in simulations are listed in Table \ref{tabl1}.
\begin{table}[t]
  \caption{Parameter values in simulations.}\label{tabl1}
        \begin{center}
        \begin{tabular}{c|c}
              \hline
              Parameter                                     & Value                                     \\\hline
              Path loss exponent $\rho$                     & $-68.5$ dB                                \\\hline
              Path loss coefficient $e$                     & $1.61$                                    \\\hline
              Distance $\varphi_0$                          & $30$ m                                    \\\hline
              Number of VUE-pair group $|\mathcal{I}|$      & $15$                                      \\\hline
              Frequency bandwidth $w$                       & $500$ kHz                                 \\\hline
              Aggregate interference $N$                    & $2 \times 10^{-12}$ W                     \\\hline
              Noise power spectral density $\sigma^2$       & $3.98 \times 10^{-21}$ W/Hz               \\\hline
              Scheduling slot duration $\delta$             & $9$ ms                                    \\\hline
              Queue termination probability $\gamma$        & $0.1$                                     \\\hline
              Weight of transmit power $\alpha_k$           & $6$, $\forall k \in \mathcal{K}$          \\\hline
              Data packet size $\mu$                        & $5$ kb                                    \\\hline
              Maximum transmit power $c^{(\max)}$           & $2$ W                                     \\
              \hline
        \end{tabular}
    \end{center}
\end{table}

For the purpose of performance comparisons, we simulate other three baseline algorithms as well, which are specified as follows.
\begin{enumerate}
  \item \emph{Channel-Aware:} A VUE-pair evaluates the need of occupying the frequency resource for packet transmissions based on the channel quality state at each scheduling slot and does not take into account the queue state.
  \item \emph{Queue-Aware:} A VUE-pair announces at each scheduling slot the preference of obtaining the frequency resource to maximize the expected long-term number of packets to be transmitted \cite{Fu09}.
  \item \emph{Random:} Implementing this algorithm, a VUE-pair randomly generates the bid of having the frequency resource at each scheduling slot, which means that the VUE-pair does not consider any dynamics from the network.
\end{enumerate}
Using a Channel-Aware algorithm or a Random algorithm, the vTx of a winning VUE-pair $k \in \mathcal{K}$ at each scheduling slot $t$ transmits a maximum possible number $D_k^{(\max), t}$ of data packets in the queue, namely,
\begin{align}
   D_k^{(\max), t} =
   \min\!\left\{q_k^t, \left\lfloor \frac{\delta \cdot w \cdot \log_2\left(1 + \frac{g_k^t \cdot c^{(\max)}}{N + w \cdot \sigma^2}\right)}
   {\mu}\right\rfloor\right\},
\end{align}
where $\lfloor \cdot \rfloor$ means the floor function.

\subsection{Convergence Property of the Proposed Algorithm}

We first examine if the stochastic behaviour of a VUE-pair converges when all VUE-pairs in the vehicular network behave according to the proposed online learning algorithm.
In the simulation, we select the number of VUE-pairs and the packet arrival rate as $K = 28$ and $\lambda = 6$, respectively.
The distance between the vTx and the vRx of each VUE-pair is fixed to be $\varphi = 26$ m, and the length limit of the queue at the vTx of a VUE-pair is assumed to be $q^{(\max)} = 5$.
Without loss of the generality, we plot the simulated variations in the post-decision state value functions of VUE-pair $1$, $\left\{\tilde{V}_1(\tilde{q}_1): \tilde{q}_1 \in \{0, 1, \cdots, 5\}\right\}$, across the time horizon in Fig. \ref{sim01}, which tells that the proposed learning algorithm converges within $600$ scheduling slots.

\begin{figure}[t]
  \centering
  \includegraphics[width=29pc]{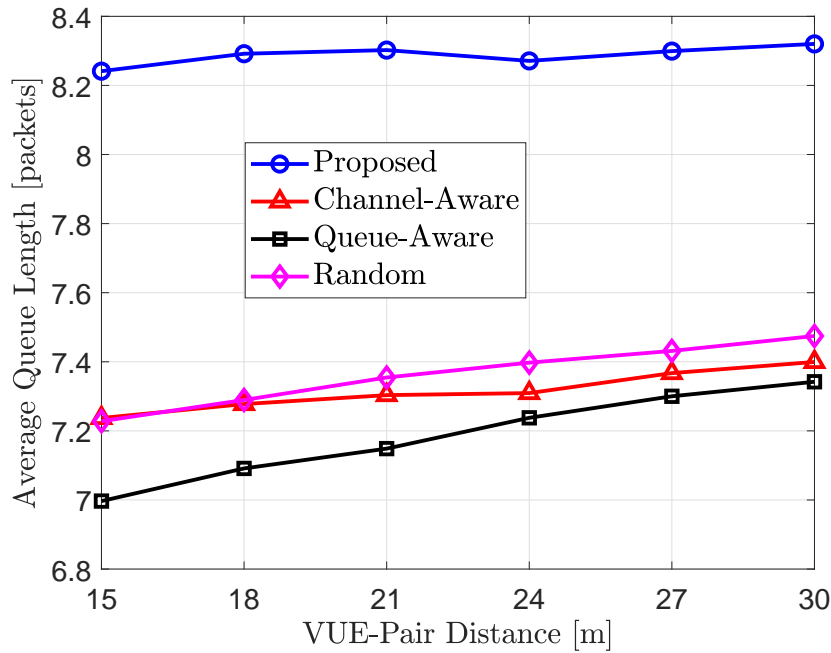}
  \caption{Average queue length per VUE-pair across the time horizon versus VUE-pair distance $\varphi$: $K = 36$ and $\lambda = 5$.}
  \label{sim2_01}
\end{figure}

\begin{figure}[t]
  \centering
  \includegraphics[width=29pc]{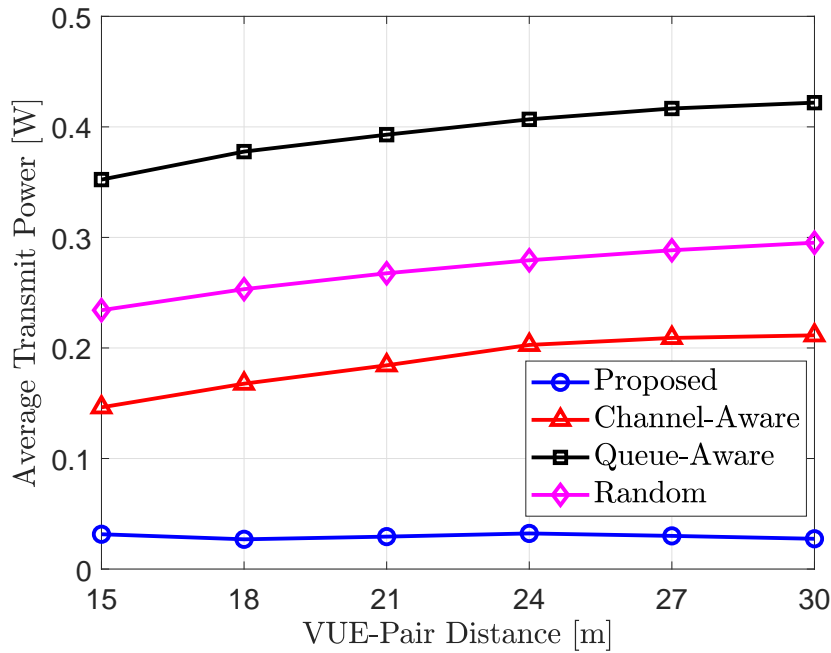}
  \caption{Average transmit power consumption per VUE-pair across the time horizon versus VUE-pair distance $\varphi$: $K = 36$ and $\lambda = 5$.}
  \label{sim2_02}
\end{figure}

\begin{figure}[t]
  \centering
  \includegraphics[width=29pc]{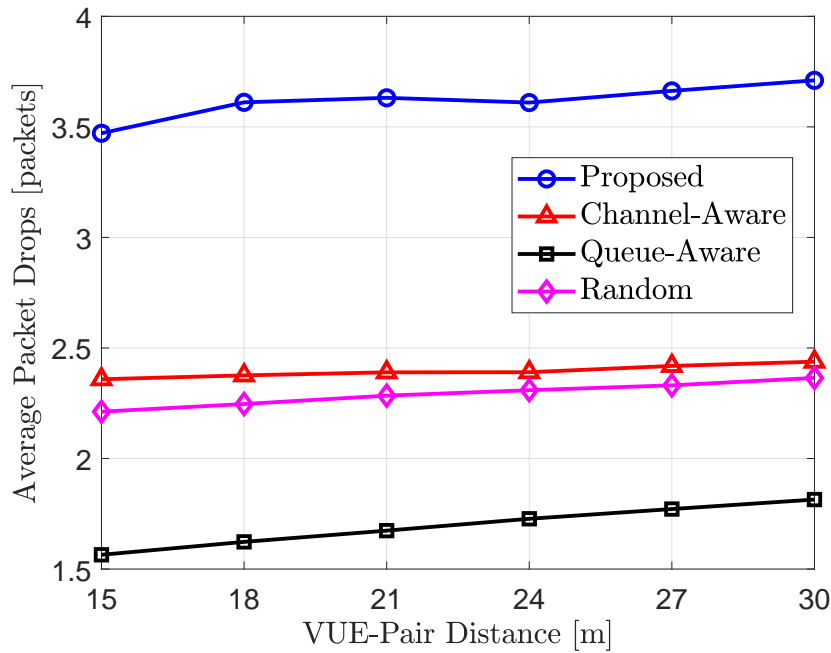}
  \caption{Average packet drops per VUE-pair across the time horizon versus VUE-pair distance $\varphi$: $K = 36$ and $\lambda = 5$.}
  \label{sim2_03}
\end{figure}

\begin{figure}[t]
  \centering
  \includegraphics[width=29pc]{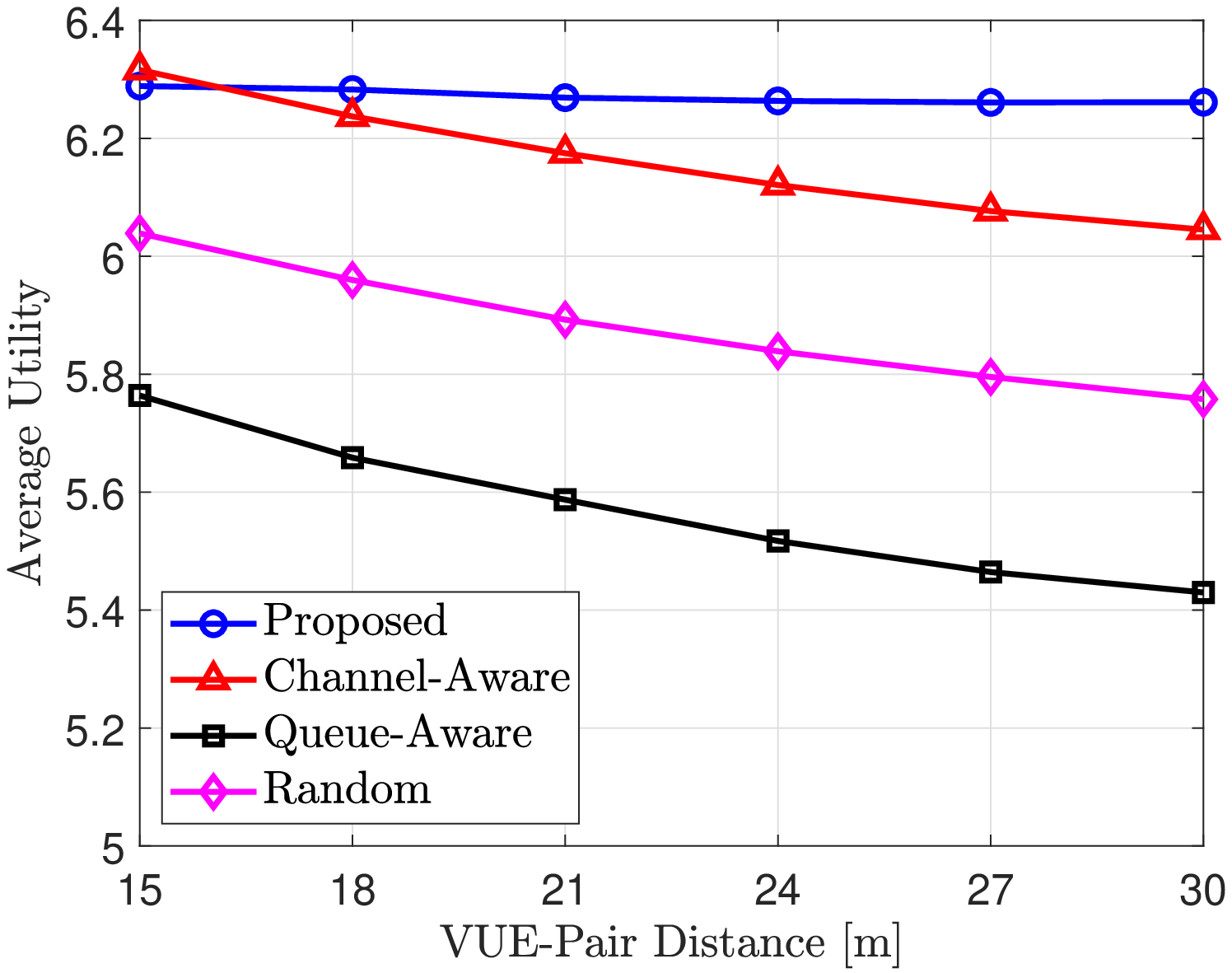}
  \caption{Average utility per VUE-pair across the time horizon versus VUE-pair distance $\varphi$: $K = 36$ and $\lambda = 5$.}
  \label{sim2_04}
\end{figure}

\subsection{Impact of $\varphi$}
\label{simu02}

Next, we demonstrate the average performance per scheduling slot in terms of the average queue length, the average transmit power consumption and the average packet drops under different values of $\varphi$.
We configure the parameter values in this simulation as: $K = 36$, $\lambda = 5$ and $q^{(\max)} = 10$.
The results are depicted in Figs. \ref{sim2_01}, \ref{sim2_02}, \ref{sim2_03} and \ref{sim2_04}.
Fig. \ref{sim2_01} illustrates the average queue length per VUE-pair per scheduling slot.
Fig. \ref{sim2_02} illustrates the average transmit power consumption per VUE-pair per scheduling slot.
Fig. \ref{sim2_03} illustrates the packet drops per VUE-pair per scheduling slot.
Fig. \ref{sim2_04} illustrates the average utility per VUE-pair per scheduling slot.

Each plot compares the performance of the proposed online learning algorithm with the other three baseline algorithms.
It can be observed from Fig. \ref{sim2_04} that the proposed algorithm achieves significant utility performance improvements when the VUE-pair distance increases to a large enough value, indicating that the proposed algorithm realizes a better trade-off between the queue length, the transmit power consumption and the packet drops.
Similar observations can be made from Fig. \ref{sim2_02}, which shows that the minimum average transmit power is consumed by the proposed algorithm.
As the distance between the vTx and the vRx of a VUE-pair increases, the channel quality becomes worse.
Hence transmitting the same number of data packets requires more power consumption, resulting in increased average queue length (as shown in Fig. \ref{sim2_01}) and average packet drops (as shown in Fig. \ref{sim2_03}).
Interestingly, by deploying the proposed algorithm, the slight increase/decrease in the transmit power consumption can be compensated by reducing/increasing the average queue length and the average packet drops.

\begin{figure}[t]
  \centering
  \includegraphics[width=29pc]{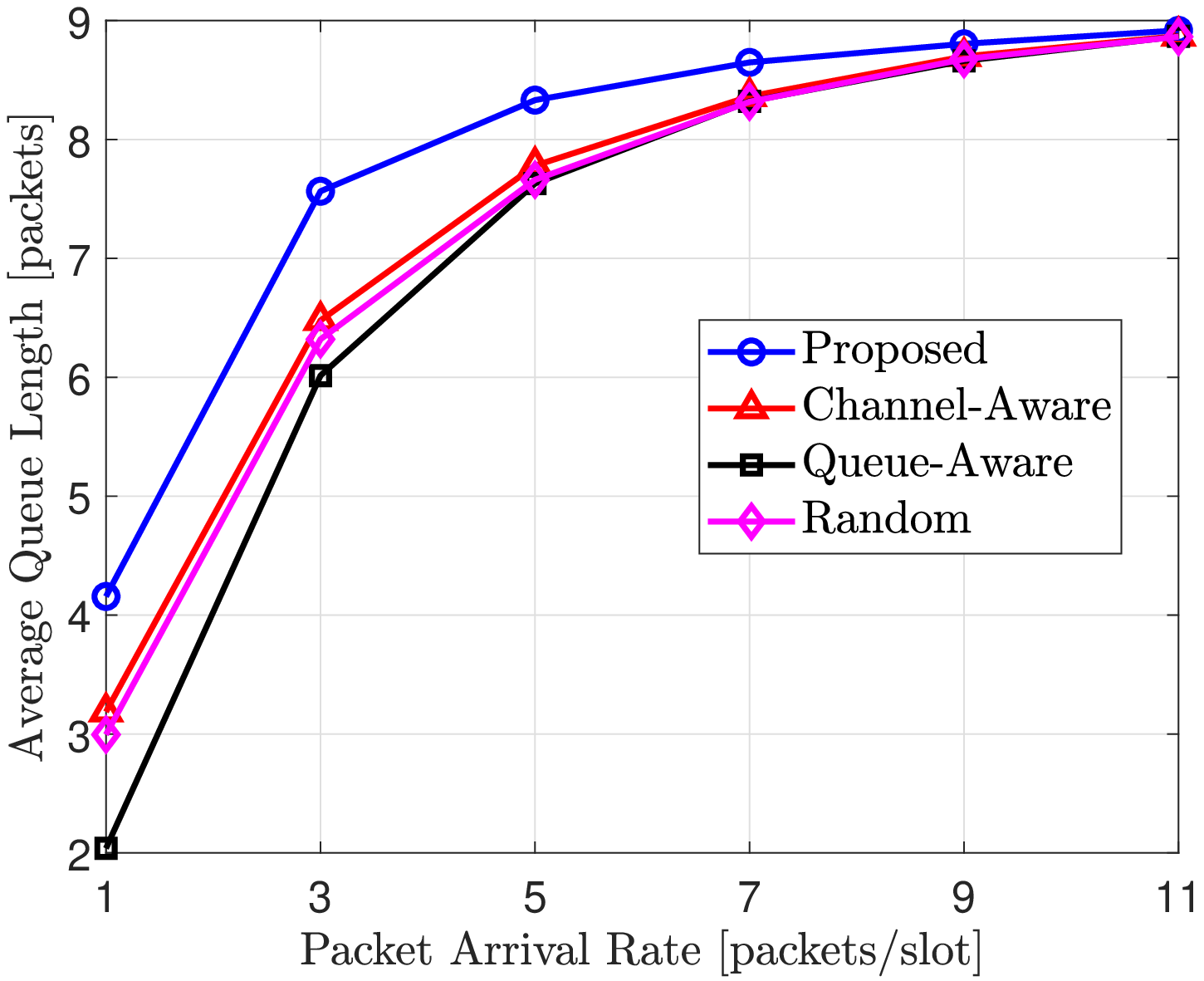}
  \caption{Average queue length per VUE-pair across the time horizon versus average packet arrival rate $\lambda$: $K = 56$ and $\varphi = 28$ m.}
  \label{sim3_01}
\end{figure}

\begin{figure}[t]
  \centering
  \includegraphics[width=29pc]{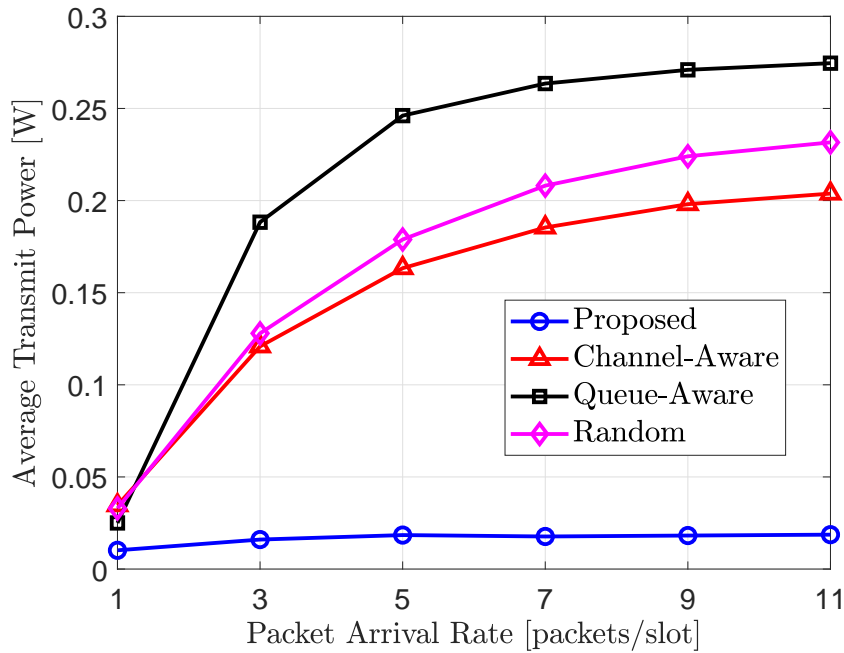}
  \caption{Average transmit power consumption per VUE-pair across the time horizon versus average packet arrival rate $\lambda$: $K = 56$ and $\varphi = 28$ m.}
  \label{sim3_02}
\end{figure}

\begin{figure}[t]
  \centering
  \includegraphics[width=29pc]{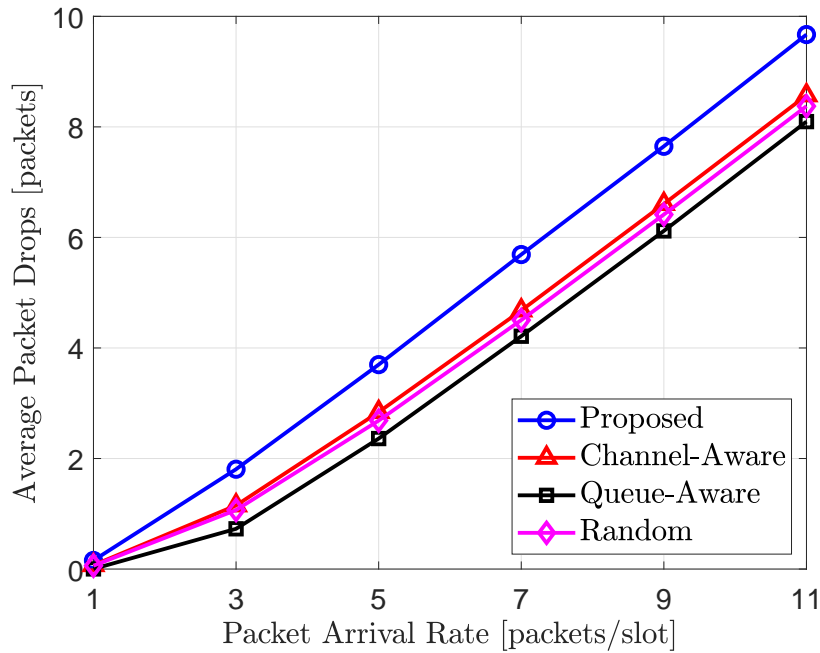}
  \caption{Average packet drops per VUE-pair across the time horizon versus average packet arrival rate $\lambda$: $K = 56$ and $\varphi = 28$ m.}
  \label{sim3_03}
\end{figure}

\begin{figure}[t]
  \centering
  \includegraphics[width=29pc]{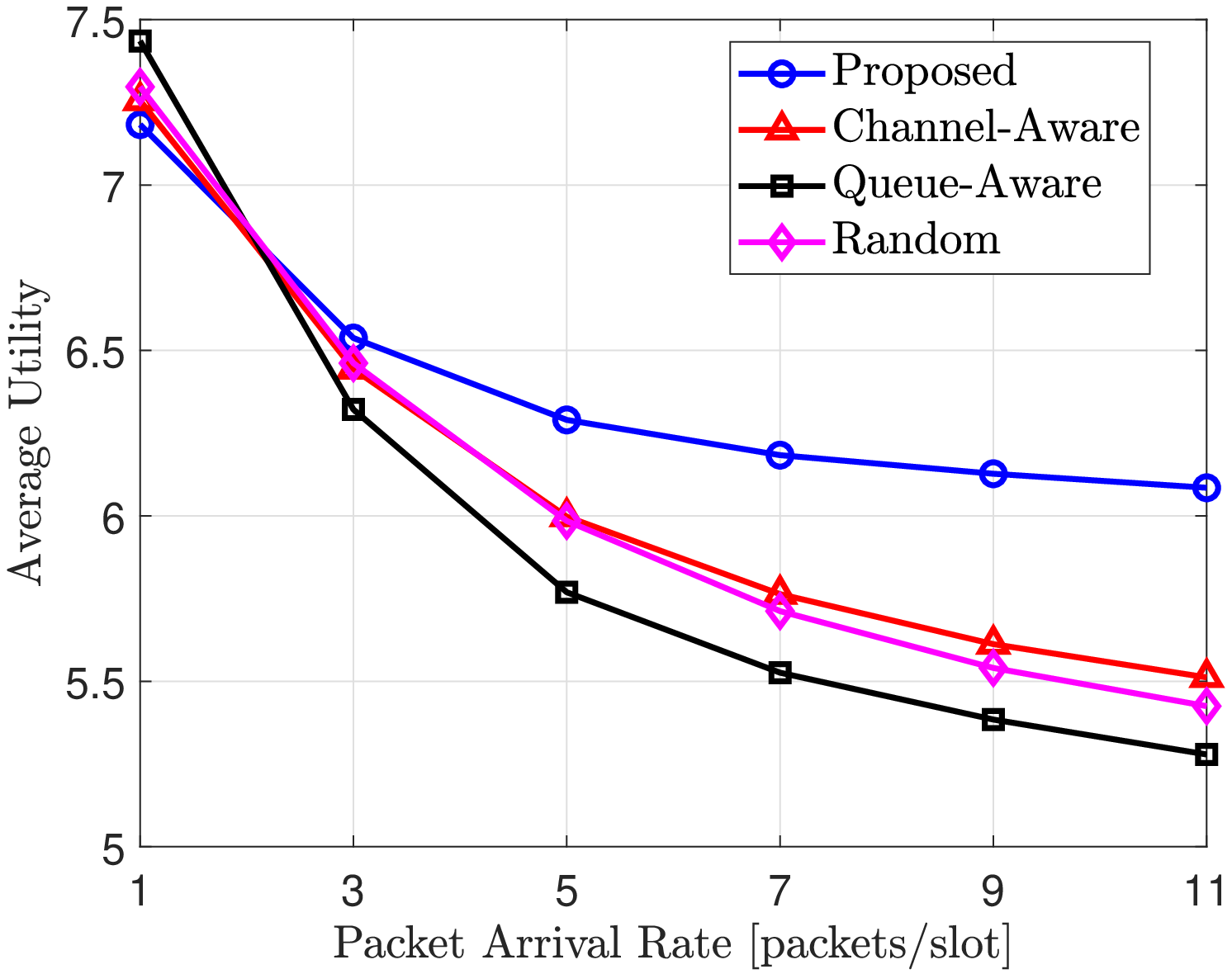}
  \caption{Average utility per VUE-pair across the time horizon versus average packet arrival rate $\lambda$: $K = 56$ and $\varphi = 28$ m.}
  \label{sim3_04}
\end{figure}

\subsection{Impact of $\lambda$}
\label{simu03}

We then exhibit the average queue length, the average transmit power consumption, the average packet drops and the average utility per VUE-pair under various settings of the average packet arrival rate in Figs. \ref{sim3_01}, \ref{sim3_02}, \ref{sim3_03} and \ref{sim3_04}.
During the simulation, we distribute $K = 56$ VUE-pairs in the network and the distance between the vTx and the vRX of a VUE-pair is fixed to be $\varphi = 28$ m.
Other parameter values are the same as the simulation in Section \ref{simu02}.
We can see from Fig. \ref{sim3_04} that as the average packet arrival rate increases, the VUE-pairs receive a smaller average utility.
With more arriving packets, the VUE-pairs consume more transmit power (as shown in Fig. \ref{sim3_03}) to deliver the queued data packets in order to avoid possible packet drops.
While the increases in the average queue length (as shown in Fig. \ref{sim3_01}) and the average packet drops (as shown in Fig. \ref{sim3_02}) are due to the maximum transmit power constraint at the vTx of a VUE-pair.
The simulations in this section and the previous Section \ref{simu02} clearly illustrate that the proposed algorithm is able to ensure better average utility performance for the VUE-pairs than the other three baseline algorithms given worse channel qualities and heavier traffic demands.

\begin{figure}[t]
  \centering
  \includegraphics[width=29pc]{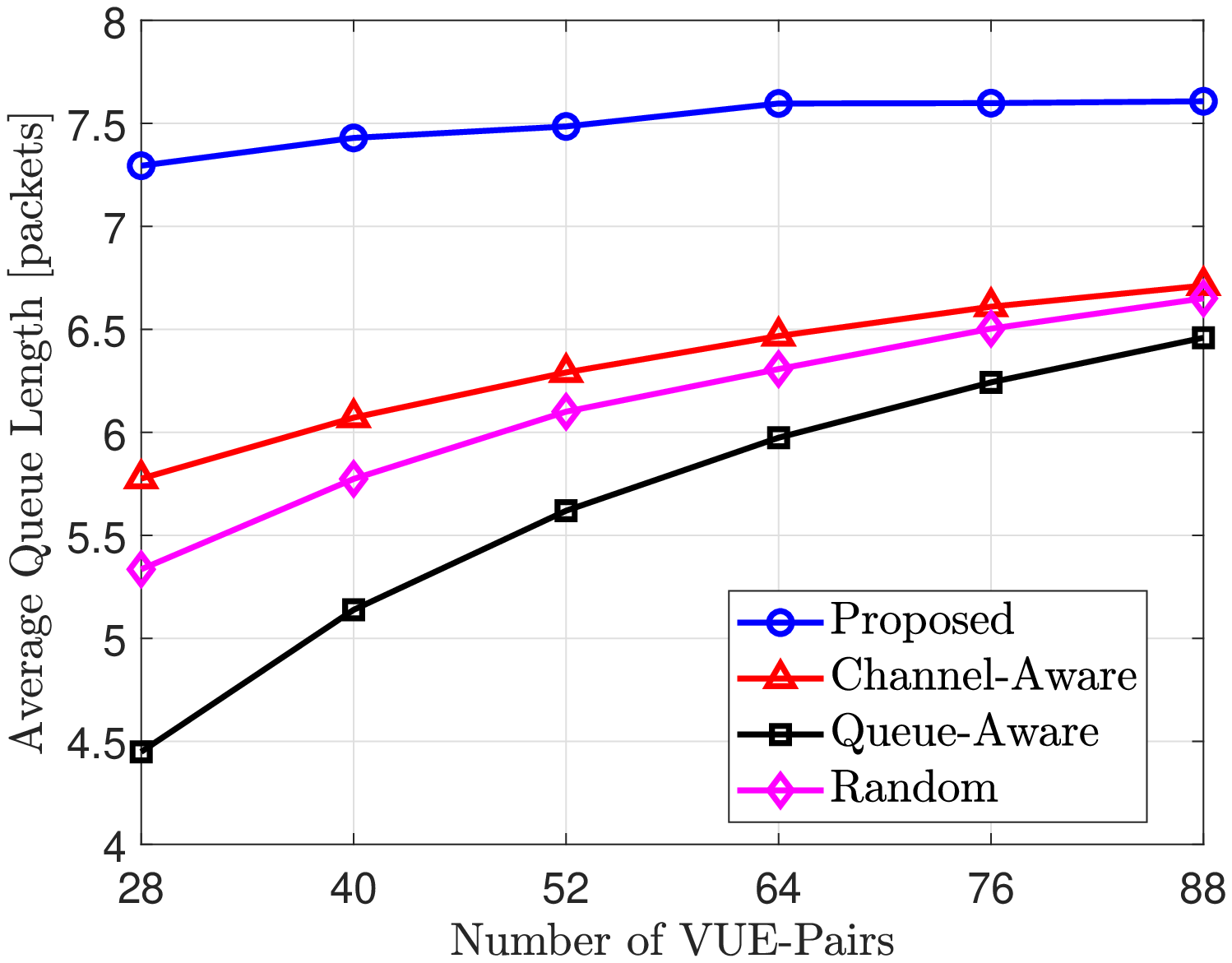}
  \caption{Average queue length per VUE-pair across the time horizon versus number of VUE-pairs $K$: $\varphi = 20$ m and $\lambda = 3$.}
  \label{sim4_01}
\end{figure}

\begin{figure}[t]
  \centering
  \includegraphics[width=29pc]{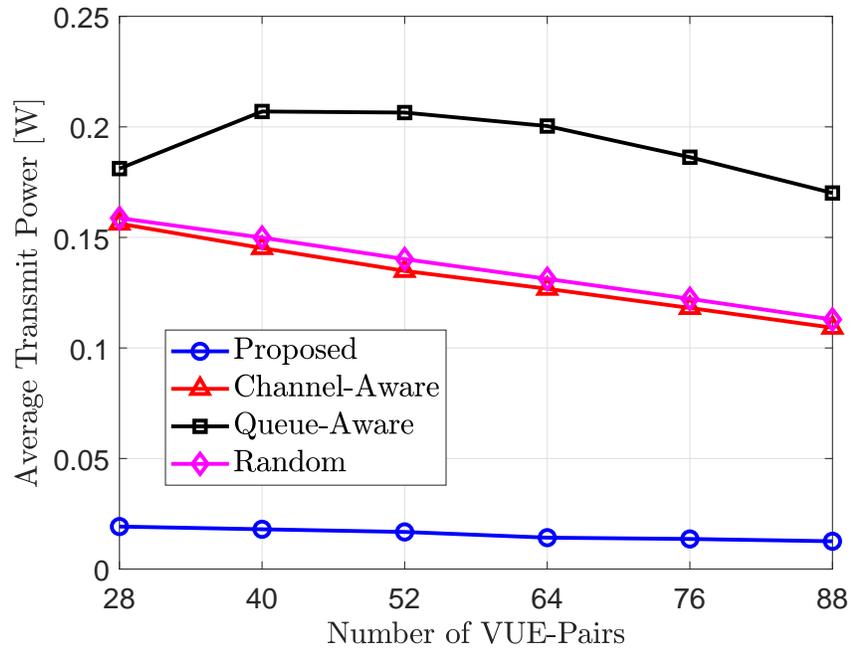}
  \caption{Average transmit power consumption per VUE-pair across the time horizon versus number of VUE-pairs $K$: $\varphi = 20$ m and $\lambda = 3$.}
  \label{sim4_02}
\end{figure}

\begin{figure}[t]
  \centering
  \includegraphics[width=29pc]{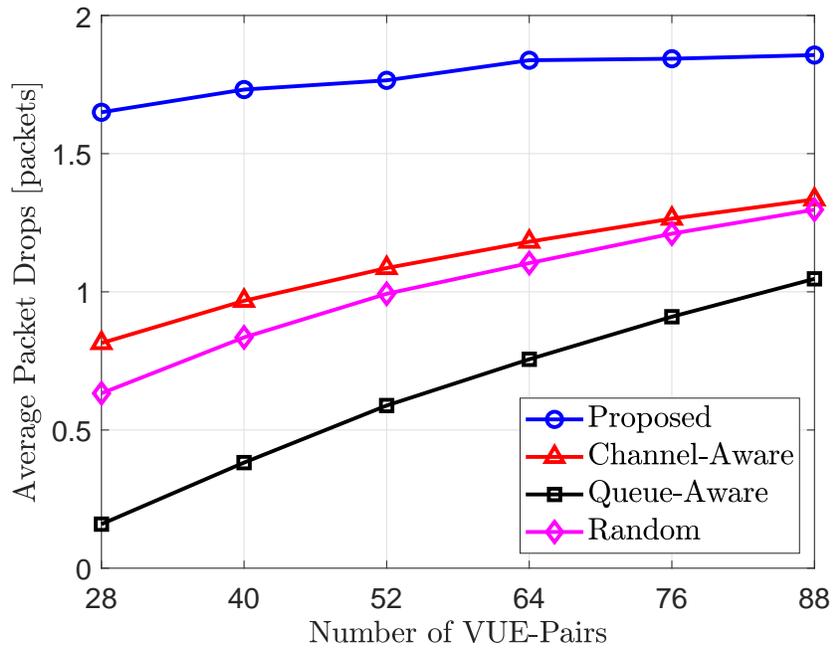}
  \caption{Average packet drops per VUE-pair across the time horizon versus number of VUE-pairs $K$: $\varphi = 20$ m and $\lambda = 3$.}
  \label{sim4_03}
\end{figure}

\begin{figure}[t]
  \centering
  \includegraphics[width=29pc]{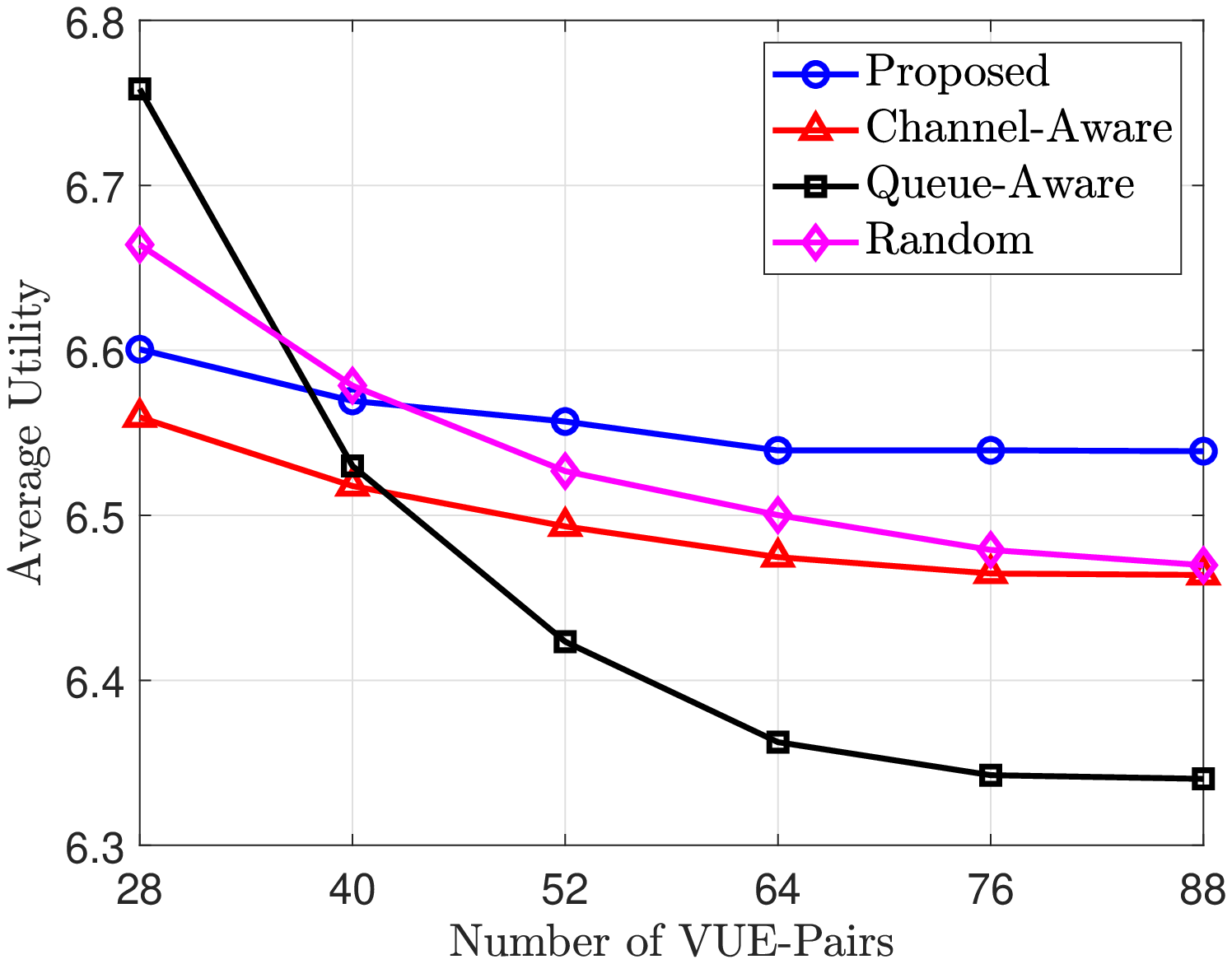}
  \caption{Average utility per VUE-pair across the time horizon versus number of VUE-pairs $K$: $\varphi = 20$ m and $\lambda = 3$.}
  \label{sim4_04}
\end{figure}

\subsection{Impact of $K$}

It becomes natural to compare the performance from the proposed algorithm to the other three baseline algorithms versus different numbers of VUE-pairs with relatively good transmission qualities between the vTxs and the vRxs and light traffic demands at the vTxs.
That is, we simulate a V2V communication network in which the VUE-pair distance and the average packet arrival rate are assumed to be $\varphi = 20$ m and $\lambda = 3$.
Figs. \ref{sim4_01}, \ref{sim4_02}, \ref{sim4_03} and \ref{sim4_04} draw the curves of simulated average queue length, average transmit power consumption, average packet drops and average utility per VUE-pair over the scheduling slots.

It is straightforward that more VUE-pairs means higher competition and hence less opportunity for occupying the single frequency resource to transmit the queued data packets, which indicates larger average queue length (as shown in Fig. \ref{sim4_01}), more average packet drops (as shown in Fig. \ref{sim4_02}), smaller average transmit power consumption (as shown in Fig. \ref{sim4_03}) and worse utility performance (as shown in Fig. \ref{sim4_04}) for all four algorithms.
Note that for the Queue-Aware algorithm, the average transmit power consumption first increases but then decreases.
This observation can be explained by the fact that with the Queue-Aware algorithm, each VUE-pair bids for the frequency resource based on the queue status.
The increase in average queue length due to higher competition pushes the VUE-pairs to more actively participate in the frequency resource auction.
More importantly, we can easily find from Fig. \ref{sim4_04} that when the number of VUE-pairs appearing in the V2V network is small, the proposed online learning algorithm does not outperform the other three baseline algorithms, which is consistent with the simulations in Section \ref{simu02} and Section \ref{simu03}.
However, when the number of VUE-pairs increases to a big enough value, a significant improvement in average utility performance can be expected from the proposed algorithm.
This trend confirms Theorem 4.

\section{Conclusions}
\label{conc}

In this paper, we investigate the problem of radio resource scheduling in a non-cooperative V2V communication network.
The VUE-pairs compete with each other for the limited frequency resource at the beginning of each scheduling slot, which is controlled by the RSU through a sealed second-price auction mechanism.
Under the assumptions of high vehicle mobility and time-varying packet arrivals, the problem is originally formulated as a stochastic game.
Using the definition of a partitioned control policy profile, the stochastic game with a semi-continuous global network state space is hence transformed into an equivalent game with a global queue state space of finite size.
The other challenge lies in the global queue state space explosion that happens in a V2V network with a large number of VUE-pairs.
Therefore, we adopt a OE to approximate the MPE, which characterizes the optimal solution to the equivalent game.
We theoretically study the AME property of a OE solution.
Without a priori statistics knowledge of queue state transitions, we derive an online algorithm to approach the OE control policy.
From numerical simulations, significant gains in utility performance from the proposed learning algorithm can be expected.

\section*{Acknowledgements}

This work was supported in part by the Finnish Funding Agency for Innovation (TEKES) under the project ``Wireless for Verticals (WIVE)''.
WIVE is a part of 5G Test Network Finland (5GTNF).

\end{document}